\documentstyle[11pt]{article}

\def\bib{\par\noindent\hangindent=3mm\hangafter=1}
\def\ApJ{Astrophys. J.}

\def\JGR{Journal Geophys. Res.}

\def\disp{\displaystyle}
\newcommand{\dpar}[2]{{\partial #1\over\partial #2}}
\newcommand{\multic}[1]{\multicolumn{1}{c}{#1}}

\input{psfig.tex}

\begin{document}

\title{\bf A Comparison of the Interiors of Jupiter and Saturn}
\author{{\sc Tristan Guillot}\\
Observatoire de la C\^ote d'Azur\\
Laboratoire G.D. Cassini, CNRS UMR 6529\\
BP 4229\\
06304 Nice Cedex 4, France\\
E-mail: guillot@obs-nice.fr
}
\date{Submitted December 12, 1998; Revised May 5, 1999}

\maketitle

\begin{center}
{\sl Planetary and Space Science}, \underline{in press}\\
Nantes Symposium special issue
\end{center}
\bigskip
\bigskip

\noindent Number of manuscript pages: 25 
Number of tables: 6 \\
Number of figures: 10 \\

\bigskip\bigskip
\noindent KEYWORDS: Jupiter, Saturn\\
\hphantom{KEYWORDS:} Interiors, planets\\
\hphantom{KEYWORDS:} Chemical abundances

\bigskip\bigskip\noindent 
RUNNING HEAD: Interiors of Jupiter and Saturn

\newpage

\begin{abstract}
\bf
Interior models of Jupiter and Saturn are calculated and compared in
the framework
of the three-layer assumption, which rely on the perception that both
planets consist of three globally homogeneous regions: a dense core, a
metallic hydrogen envelope, and a molecular hydrogen envelope.
Within this framework, constraints on the core mass and abundance of heavy
elements (i.e. elements other than hydrogen and helium) are given by
accounting for uncertainties on the measured gravitational moments,
surface temperature, surface helium abundance, and on the inferred
protosolar helium abundance, equations of state, temperature
profile and solid/differential interior rotation. 

Results obtained solely from static models matching the measured
gravitational fields indicate that the mass of Jupiter's dense core is
less than $14\rm\,M_\oplus$ (Earth masses), but that models with no core
are possible given the current uncertainties on the hydrogen-helium
equation of state. Similarly, Saturn's core mass is less than
$22\rm\,M_\oplus$ but no lower limit can be inferred. The total mass of
heavy elements (including that in the core) is constrained to lie
between $11$ and $42\rm\,M_\oplus$ in Jupiter, and $19$ to
$31\rm\,M_\oplus$ in Saturn. The enrichment in heavy elements of their
{\it molecular} envelopes is $1$ to $6.5$, and $0.5$ to $12$ times the solar
value, respectively. 

Additional constraints from evolution models accounting for the
progressive differentiation of helium (Hubbard et al. 1999) are used
to obtain tighter,
albeit less robust, constraints. The resulting core masses are then
expected to be in the range $0$ to $10\rm\,M_\oplus$, and $6$ to
$17\rm\,M_\oplus$
for Jupiter and Saturn, respectively. Furthermore, it is shown that
Saturn's atmospheric helium mass mixing ratio, as derived from
Voyager, $Y=0.06\pm 0.05$, is
probably too low. Static and evolution models favor a value of
$Y=0.11-0.25$. 
Using, $Y=0.16\pm 0.05$, Saturn's molecular region is found to be
enriched in heavy elements by $3.5$ to $10$ times the 
solar value, in relatively good agreement with the measured methane
abundance. 

Finally, in all cases, the gravitational moment $J_6$ of models
matching all the constraints are found to lie 
between $0.35$ and $0.38\times 10^{-4}$ for
Jupiter, and between $0.90$ and $0.98\times 10^{-4}$ for Saturn, assuming
solid rotation. For comparison, the uncertainties on the measured
$J_6$ are about 10 times larger. More accurate measurements of $J_6$
(as expected from the Cassini orbiter for Saturn) will therefore
permit to test the validity of interior models calculations and the
magnitude of differential rotation in the planetary interior.

\end{abstract}

\newpage

\section{Introduction}

Jupiter and Saturn played a key role in the formation of the Solar
System. Yet their internal structure and composition are still poorly
known. This is due to the fact that the only way to probe the deep
regions of these planets is by calculating interior models matching
their observed gravitational field, a method that can
only yield information on quantities that are averaged over a
significant fraction of the planetary radius. It is however of
primordial importance to derive the largest ensemble of theoretical
models compatible with observations, and yet to try to narrow down the
ensemble of solutions as much as possible, thus providing related
scientific fields with useful constraints. 

Previous work on the subject, including those of Hubbard \& Marley
(1989), Zharkov \& Gudkova (1992), Chabrier et al. (1992) and Guillot
et al. (1994b), has generally aimed at finding a sample of models
matching the observational constraints, i.e. radius and
gravitational moments. Using the new H/He mixing ratio from Galileo
and taking advantage of the improvements in computing capabilities
over the past years, Guillot et al. (1997) have
calculated an extended ensemble of interior models of Jupiter matching
all observational constraints within the error bars. 

Although new observations of Saturn relevant to studies of its
internal structure will probably have to await the arrival of the
Cassini-Huygens space mission in 2004, the present article is
motivated by the renewed interest in giant planet formation subsequent
to the discovery of extrasolar planets (see e.g. Mayor \& Queloz 1995;
Marcy and Butler 1998). As a consequence, the formation of giant planets
by gas instability (Cameron 1978), which had been abandoned to the
profit of the nucleated instability model (Lissauer 1993, Pollack et
al. 1996), has reborn from its ashes, and been proposed
as a mechanism that led to the rapid formation of Jupiter and
Saturn (Boss 1998). This prompts for a recalculation of models of
Saturn aimed at constraining its internal composition. 

Furthermore, accurate evolution calculations (Guillot et al. 1995)
using improved atmospheric models (Marley et al. 1996; Burrows et
al. 1997) were available, but assumed a homogeneous structure. The
additional energy provided by helium sedimentation (or that of any
other element) has been estimated by Hubbard et al. (1999). The resulting
model ages therefore provide an additional constraint, as they must,
within the uncertainties, agree with the age of the Solar System.

The aim of the present article is thus to calculate new static models of
Saturn, to include additional constraints obtained from evolution
calculations on the models of Jupiter and
Saturn, and to compare the structure and composition of the two giant
planets. Relevant observations and input physics are presented in
Section~2. The main assumption of a three layer structure for Jupiter
and Saturn is discussed in Section~3. I then present in the next
section the method used to calculate interior models. The results are
detailed in Section~5; their consequences on models of the
formation of the giant planets and perspectives are discussed in
Section~6.

\section{Input data}

\subsection{Gravitational field}

Jupiter and Saturn are rapid rotators: the rotation rates derived from
their magnetic fields are 9 hours and 55 minutes, and 10 hours and 39
minutes, respectively. Although the consequences of rotation on
their very structure, and in particular the presence and strength of
convection in their interiors, are difficult to assess (Stevenson and
Salpeter 1977;
see also Busse 1976, Zhang \& Schubert 1996), this property is a
chance because it allows us
to constrain their interior density profiles, by measuring the
departures from sphericity of their gravitational potentials:
\begin{equation}
U(r,\theta)={GM\over r}\left\{1-\sum_{i=1}^{\infty}\left({R_{\rm
eq}\over r}\right)^{2i} J_{2i} P_{\rm 2i}(\cos \theta)\right\},
\end{equation}
where $G$ is the gravitational constant, $M$ the mass of the planet,
$R_{\rm eq}$ its equatorial radius, $r$ the distance to the planetary
center, $\theta$ the polar angle (to the rotation axis), and 
$P_{\rm 2i}$ are Legendre polynomials. The $J_{2i}$ are the
gravitational moments. Their observed values (inferred mostly from the
trajectories of the Pioneer and Voyager spacecrafts) are given in
Table~\ref{tab:grav}. The gravitational moments can also be related to
the internal density
profile $\rho(r)$ (and thus to theoretical models) by the following
relation (e.g., Zharkov \& Trubitsyn 1974): 
\begin{equation}
J_{2i}=-{1\over M R_{\rm eq}^{2i}}\int \rho(r)r^{2i}P_{2i}(\cos
\theta)d\tau, 
\label{eq:j2i}
\end{equation}
in which the integral is calculated over the total volume of the
planet. A common method for the solution of Eq.~\ref{eq:j2i} is the
so-called theory of figures, which will not be developed here (see
Zharkov \& Trubitsyn 1978). The present work uses equations to the
third order of the theory of figures. 

\begin{table}[bth]
\begin{center}
\caption{Characteristics of the gravitational fields}
\label{tab:grav}
\begin{tabular}{l*{4}{r@{.}l}}\hline
 & \multicolumn{4}{c}{Jupiter} & \multicolumn{4}{c}{Saturn} \\
 & \multicolumn{2}{c}{measured$^{\rm a}$} &
 \multicolumn{2}{c}{adjusted$^{\rm c}$} 
 & \multicolumn{2}{c}{measured$^{\rm b}$} &
 \multicolumn{2}{c}{adjusted$^{\rm c}$} \\
 \hline
$M\rm/M_\oplus$ & 317&83 & \multicolumn{2}{c}{ } & 95&147
 &\multicolumn{2}{c}{ } \\ 
$R_{\rm eq}/10^9$ & 7&1492(4) & \multicolumn{2}{c}{ }& 6&0268(4) &
 \multicolumn{2}{c}{ }\\ 
$J_2/10^{-2}$ & 1&4697(1) & 1&4682(1) & 1&6332(10) & 1&6252(10) \\
$J_4/10^{-4}$ & $-$5&84(5) & $-$5&80(5) & $-$9&19(40) & $-$8&99(40)  \\
$J_6/10^{-4}$ & 0&31(20) & 0&30(20) & 1&04(50) & 0&94(50) \\ \hline
\end{tabular}
\noindent
\parbox{11cm}{\ {\it Note.} The numbers in parentheses are the uncertainty in
the last digits of the given value. All the quantities are relative to
the 1 bar pressure level.\\
$^{\rm a}$ Campbell \& Synnott (1985).\\
$^{\rm b}$ Campbell \& Anderson (1989).\\
$^{\rm c}$ Adjusted for differential rotation using Hubbard (1982).}
\end{center}
\end{table}

A complication arises from the fact that the equations derived from
that theory generally assume the planet to be rotating as a solid body. 
Observations of the atmospheric winds show significant variations with
latitude, however (e.g., Gierasch \& Conrath 1993). The question of the
depth to which these differential
rotation patterns extend is still open. Hubbard (1982) has proposed
a solution to the planetary figure problem in the case of a deep rotation
field that possesses cylindrical symmetry. It is thus possible to
derive, from interior models assuming solid rotation, the value of the
gravitational moments that the planet would have if its surface
rotation pattern extended deep into its interior. It is {\it a priori}
impossible to prefer one model to the other, and I will therefore
present calculations assuming both solid and differential rotation. 
Table~\ref{tab:grav} gives both the measured gravitational
moments, and those corrected for differential rotation.

\subsection{Atmospheric abundances}

The following definitions for chemical abundances will be used:
$X$, $Y$ and $Z$ are the mass mixing
ratios of hydrogen, helium, and heavy elements, respectively. Because
it is difficult to separate helium and heavy elements, I 
also introduce an {\it equivalent\/} helium mass mixing
ratio, $Y_Z\sim Y+Z$ (see Eq.~\ref{eq:Z} hereafter for an exact
derivation). 

A first indicator of the internal evolution of the giant planets is
the abundance of helium in their atmospheres. Since the helium to
hydrogen ratio in the whole planet must have remained constant for its
entire evolution, any measured depletion in the atmosphere compared to
the initial value means that more helium is present deeper inside the
planet. This has significant consequences on the planet's internal structure
and evolution. Unfortunately, the initial (i.e., protosolar) helium
mass mixing ratio 
cannot be directly inferred from abundance measurements in the solar
photosphere because of the progressive gravitational settling of helium into the
radiative zone of our star. Instead, determination of the protosolar
value must rely on solar evolution models including 
diffusion and satisfaction of the constraints provided by observations of solar
oscillations. These indicate that the protosolar helium mass mixing
ratio was $Y/(X+Y)=0.270\pm 0.005$ (Bahcall \& Pinsonneault 1995). 

In situ measurements of that quantity in Jupiter yield $Y/(X+Y)=0.238\pm
0.007$ (von Zahn et al. 1998). Combined radio occultation
measurements and spectra analysis indicate that, in Saturn, $Y/(X+Y)=0.06\pm
0.05$ (Conrath et al. 1984). However, similar measurements for Jupiter
led to $Y/(X+Y)=0.18\pm 0.04$ (Gautier et al. 1982), in 
disagreement with the value obtained by the Galileo probe. It is
conceivable that the value obtained for Saturn is incorrect too,
prompting us to consider other values for the atmospheric helium
mixing ratio in that planet. Indeed, both a reexamination of Voyager
IRIS data (Gautier \& Conrath, personal communication 1998) and
constraints from the static and evolutionary models (see below, and
accompanying paper by Hubbard et al. 1999) point to significantly higher
values of $Y$.
In any case, the conclusion that more helium was present in the
protosolar nebula gas from which Jupiter and Saturn formed than is
observed today in their atmospheres seems inescapable. This is
explained by the presence of either a first order molecular/metallic
phase transition of hydrogen, or a hydrogen/helium phase separation,
or both, as described below. 

The measured abundances of other elements also provide important
clues to the composition of the planets. Both Jupiter and Saturn are
globally enriched in heavy elements compared to the Sun.
In Jupiter, the {\it in situ}
measurements of the Galileo probe are compatible with a $\sim$3 times
solar enrichment of carbon and sulfur (Niemann et al. 1998). It is still
unclear as to whether nitrogen is close to solar (by a
factor 1 to 1.5; de Pater and Massie 1985), moderately (2.2 to 2.4;
Carlson et al. 1992) or strongly enriched (3.5 to 4.5 times solar;
Folkner et al. 1998). Water is still a problem because of its
condensation at deep levels, and only a lower limit of $\sim$0.1 times
solar can be inferred from the measurements. On the contrary, neon,
and perhaps argon appear to be depleted, having abundances less than
0.13 and 1.7 times solar, respectively (Niemann et al. 1998). This
might be due to the fact
that noble gases, in particular neon, tend to be carried deep into the
interior by helium-rich droplets (Roulston \& Stevenson 1995).

Unfortunately, the uncertainties for Saturn are still relatively
large. Its atmosphere is enhanced in carbon by a factor of 2 to 7, and
in nitrogen by a factor 2 or more (Gautier \& Owen
1989). Observationally, it could therefore be more rich in heavy
elements than the jovian atmosphere. This will be tested by the
Cassini-Huygens mission.

\subsection{Atmospheric temperatures}

The temperatures at the tropopause (at pressures of about 0.3 bar) are
relatively well constrained by direct inversions of infrared
spectra. These predict relatively large latitudinal temperature changes
of the order of 10 K (Conrath et al. 1989). The temperature gradients
decrease with
tropospheric depth, as interior convection presumably becomes more
efficient in redistributing the heat. 
However, the accuracy of this method drops rapidly with
increasing pressure and does not reach levels deep
enough to be used as surface condition for interior models. So far,
the only reliable measurement of the deep tropospheric temperature of
a giant planet is that from the Galileo probe in Jupiter: 166 K at 1
bar (Seiff et al. 1998). It is not clear however how representative of
the whole planet this measurement is. Previous analyses have relied
upon (local) radio occultation data acquired with the Pioneer and Voyager
spacecrafts (Lindal et al. 1981, 1985) that predicted 1 bar
temperatures of $165\pm 5$ K in Jupiter and $134.8\pm 5$K in
Saturn. The temperatures inferred from these data are however dependent
on the assumed mean molecular weight $\overline{m}$. In the case of a
vertically uniform atmospheric composition (a fairly accurate
representation in the present
case), one can derive the temperature directly from the measured
refractivity $N$ as a function of altitude $z$:
\begin{equation}
T(z)=-{\overline{m}\over k_{\rm B}N(z)}\int_\infty^z N(h)g(h)dh,
\label{eq:tz}
\end{equation}
where $g$ is the gravity and $k_{\rm B}$ is the Boltzmann constant. 

The new measurement of the helium mixing ratio in Jupiter yields a
higher $\overline{m}$, and hence a larger surface temperature of 170.4
K at 1 bar. On the other hand, the
interpretation of the IRIS data suggests that the radio occultation
data tend to predict temperatures that are too high by a few kelvins 
(Gautier, personal communication 1998),
therefore pointing to a value closer to 165 K. In this work, I choose
to consider values of 165 to 170 K for Jupiter.

In the case of Saturn, I will show that values of $Y$ higher than the
Voyager value ($Y=0.06$), are more in agreement with the evolution of
this planet, and that the surface helium mass
mixing ratio could be as high as 0.25. According to Eq.~\ref{eq:tz},
these high values indicate that
the planet's surface temperature might be significantly
underestimated, and be of 
the order of 145 K, a value that I choose to use in the
calculations. I also present calculations with 1 bar temperatures of
135 K. The lower limit of 130 K appears to be inconsistent with the
static and evolutionary constraints and is therefore not used.

\subsection{Equations of state}

Inside Jupiter and Saturn, all chemical species are compressed to very
high pressures (up to 10-70 Mbar) at temperatures for which
experimental data are difficult to obtain, and theoretical models very
complex. Hydrogen undergoes a
transition from a molecular phase at low pressures to a metallic phase
at pressures larger than about one Mbar. Yet it is still unclear
whether this transition is discontinuous (first order) or continuous. 

To date, the most thorough theoretical
effort to model the high-pressure, low-temperature liquid hydrogen
equation of state (EOS) predicts that this element passes from
molecular to metallic phase via a first order (i.e. discontinuous)
plasma phase transition (PPT) (Saumon et al. 1995). This EOS is
however based on the so-called ``chemical picture'' which assumes that
the species considered (H, H$_2$, H$^+$, H$^-$) are
chemically distinct under all conditions. This assumption is probably
not entirely satisfied in the region of interest, and hence, Saumon et
al. (1995) provide two EOSs: a thermodynamically consistent one, which
includes the PPT, and one that has been smoothly interpolated between
low-pressure and high-pressure regimes. These two EOSs (referred to as
``PPT-EOS'' and ``$i$-EOS'' hereafter) are used in
this work and serve as an estimate of present uncertainties on
the calculated densities of hydrogen at high pressure.

Recent shock compression on liquid deuterium (e.g., Holmes et al.
1995; Weir et al. 1996; Collins et al. 1998) show that the
molecular/metallic transition is 
even more complex than was expected. At pressures of about 1.4
Mbar, the resistivity of the 
liquid ceases to decrease, which is interpreted as the sign that the
metal phase has been reached, thus favoring the case for a continuous
molecular to metallic transition (Nellis et al. 1999; Da Silva et
al. 1997). However, the
measured resistivity is still larger than theoretical estimates
pertaining to a fully-ionized metal (Hubbard et al. 1997). 
As discussed by Nellis et al. (1998), in the $1-2$ Mbar pressure range,
the transition is from a semi-conducting to an essentially metallic
fluid but one which retains a strong pairing character. As these
authors note, ``the precise mechanism by which a metallic state might
be attained is still a matter of debate''. One proposition is that 
the PPT exists, but lies deeper still (Saumon et al. 1999). The
complex behavior of hydrogen due to the molecular/metallic transition
probably extends over the range $1-3$ Mbar indicated in
Fig.~\ref{fig:jupsat} hereafter.

Helium is the source of yet another difficulty: its behavior would be
simpler than that of hydrogen, since it requires larger pressures to become
ionized, were it not for its interactions with hydrogen atoms and
molecules, and the subsequent possibility that a phase {\it
separation} between helium-rich and helium-poor mixtures occurs in
Jupiter and Saturn (Salpeter 1973). To date, no fully consistent
equation of state including both hydrogen and helium has been made
available. Attempts to predict the critical pressures and temperatures
at which helium separates from hydrogen still yield very different
results (Pfaffenzeller et al. 1995; Klepeis et al. 1991) but generally
agree that this should take place in the metallic region (see also
Stevenson 1982). I choose to consider this phase separation as a
possibility, but do not attempt to derive a consistent
phase-diagram. Instead, in the framework of the Saumon-Chabrier EOS,
helium is added to hydrogen using the ideal mixing rule (thereby
preventing any derivation of Gibbs' mixing enthalpy). 

Heavier elements are present in smaller quantities, so that
errors on their EOSs should not lead to substantial changes in the
final interior models. Their contribution is derived from the ideal
mixing rule, with the assumption that their EOS is proportional to
that of helium, with a factor equal to the ratio between the molecular
mass of helium and heavy elements. However, contrary to previous work
(see Chabrier et al. 1992, Guillot et al. 1994b, 1997), I account for
their specific heat, as it is significantly different from that of
helium. The additional entropy due to the presence of heavy elements
is derived within the perfect gas formalism. This is clearly an
oversimplification, but this term is only significant in the molecular
phase and not in the partly degenerate metallic phase, for which
density is less sensitive to temperature changes. 
Accordingly, the specific entropy of the gas at any
given temperature and pressure is written:
\begin{equation}
S(T,P,Y,Z)=X S_{\rm H}(P,T) + Y S_{\rm He}(P,T) + S_{\rm mix}(P,T,Y) +
Z{k_{\rm B}\over m_z} \left[c_{p\,z}\ln {T\over T_0} - \ln {P\over
P_0}\right],
\label{eq:S}
\end{equation}
where $S_{\rm H}$ and $S_{\rm He}$ are the entropies for pure hydrogen
and helium, respectively, $S_{\rm mix}$ is the entropy of mixture of
these two species (see Saumon et al. 1995), $m_z$ is the (mean)
molecular mass of the heavy elements, $c_{p\,z}k_{\rm B}$ their
specific heat (in units of $\rm erg\, K^{-1}\,g^{-1}$) and $T_0$ and
$P_0$ are arbitrary reference temperature and pressure, respectively. 
I use $c_{p\,z}=4$ and $m_z=17$ as typical values. Different values
were tested without significantly changing the results. In fact, most
of the difference with previous calculations is the result of the use
of $Y$ 
(i.e. in the molecular envelope, the observed helium mass mixing
ratio) instead of the equivalent helium mass fraction $Y_Z$ in
Eq.~\ref{eq:S} (see Guillot et al. 1994b). Note that in the molecular
envelope, in which the temperature profile is most important, helium
is in atomic form and has the smallest possible specific heat. 
The use of Eq.~\ref{eq:S} is thus especially important for Saturn,
whose molecular envelope is rich in heavy elements. The resulting 
models of this planet are substantially cooler than
previous ones. 

In the central cores of the planets, I use the density relations
appropriate to ``ices'' (a mixture of CH$_4$, NH$_3$, and H$_2$O)
and ``rocks'' (Fe, Ni, MgO and SiO$_2$), as described by Hubbard \&
Marley (1989).

\begin{figure}[ht]
\begin{center}
\psfig{figure=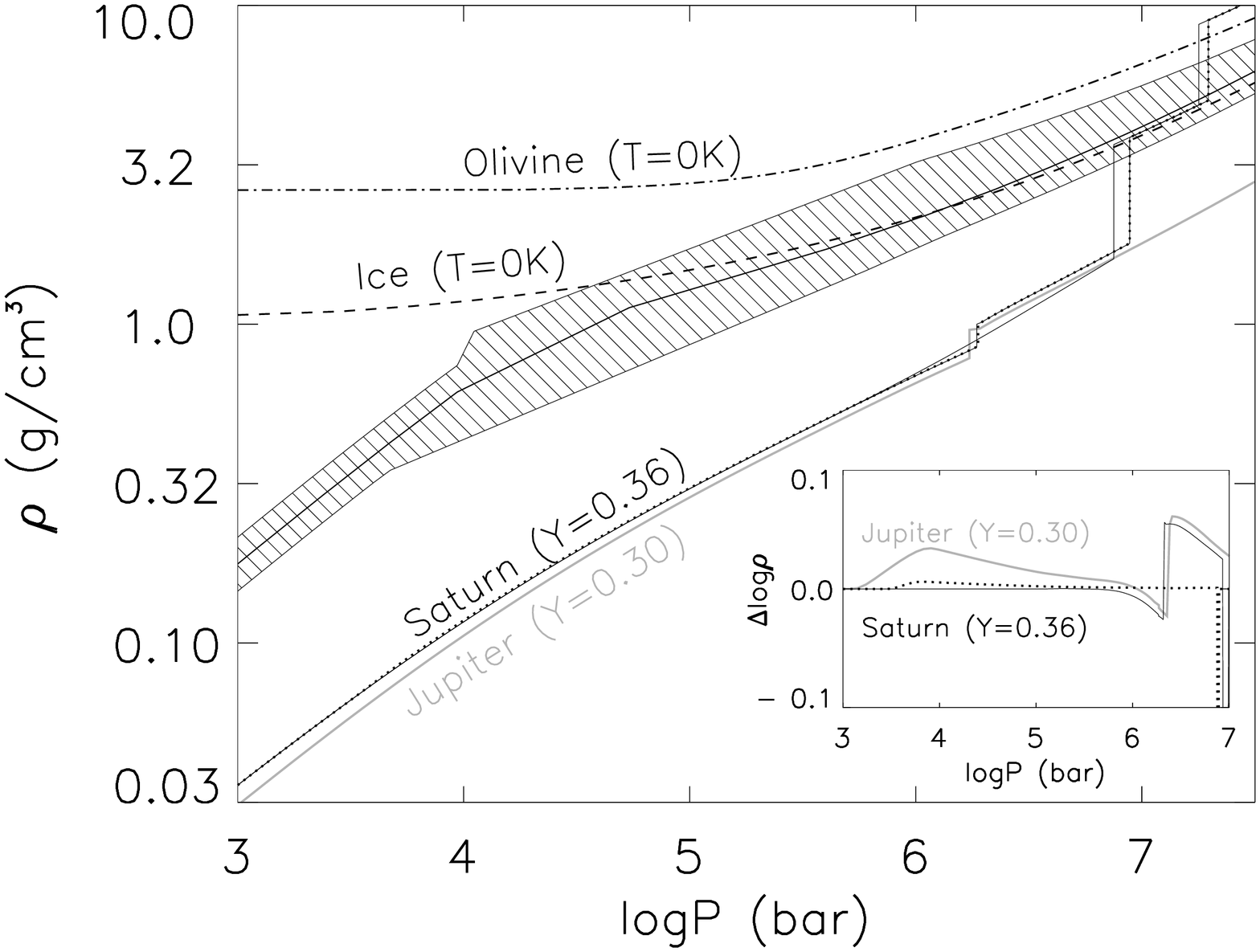,angle=90,width=16cm}
\caption{Density profiles in models of Jupiter (gray line) and Saturn
(continuous lines: adiabatic i-EOS and PPT-EOS models; dashed:
non-adiabatic i-EOS model). Upper curves (dashed and dot-dashed) are
$T=0$ K density profiles for water ice and olivine (from Thompson
1990). The dashed region represents the assumed uncertainty on the EOS
for heavy elements ($\rho_Z(P,T)$). Within this region, the continuous
line corresponds to our ``preferred'' profile for $\rho_Z$. {\it
Inset\/}: Differences of the decimal logarithm of the Saturn density
profiles with the same profile using the $i$-EOS and an adiabatic
structure (plain and dotted lines). The gray line corresponds to the
same difference but for a PPT-EOS non-adiabatic Jupiter
model. (Adapted from Guillot et al. 1997).}
\label{fig:rho_z}
\end{center}
\end{figure}

\begin{figure}[ht]
\begin{center}
\psfig{figure=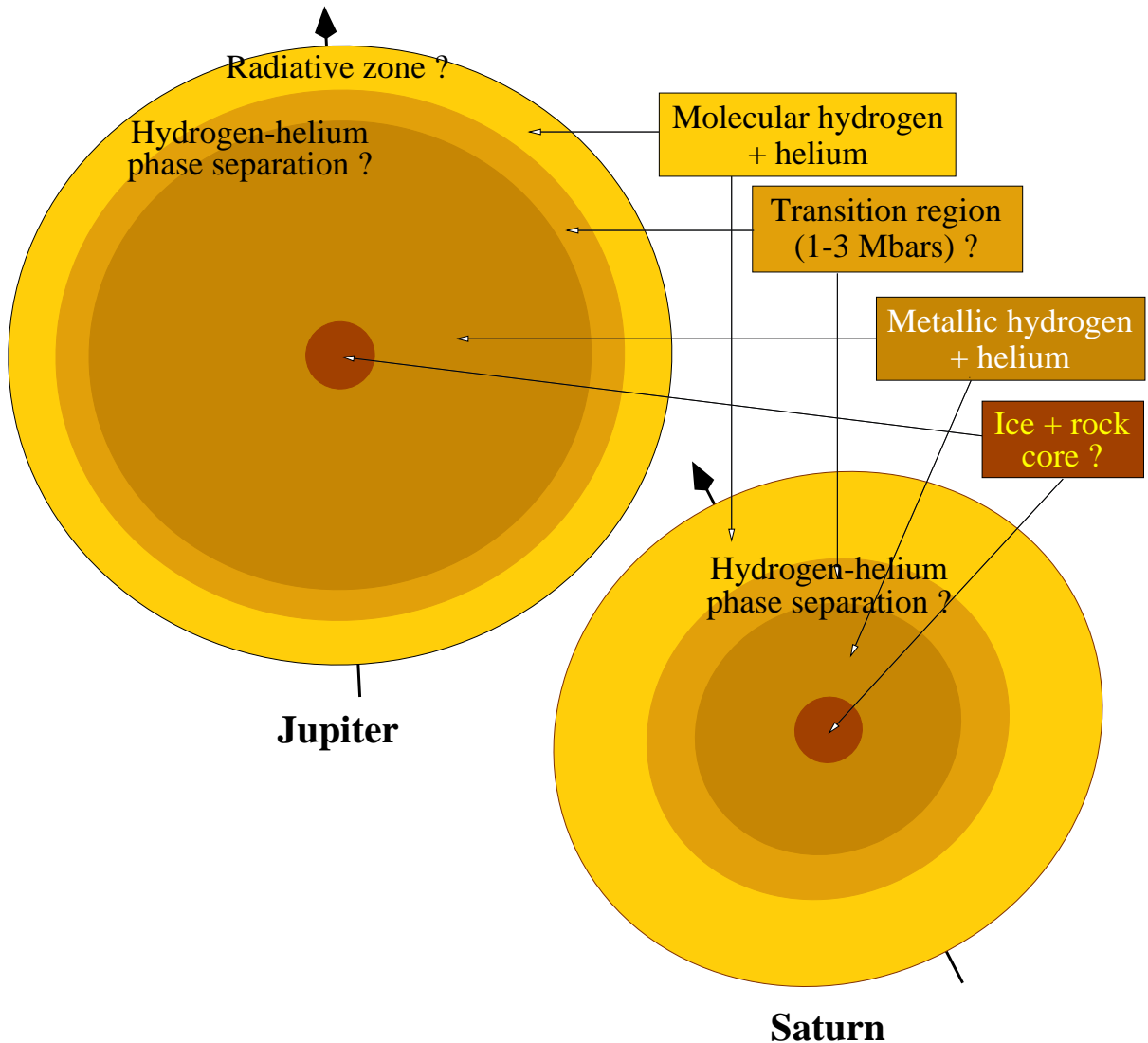,width=16cm}
\caption{Conventional view of the interiors of Jupiter and Saturn,
with a three-layer structure including the ice/rock core, the
molecular and metallic regions. A transition region, assumed to lie
between 1 and 3 Mbar, is represented, but experiments and theory are
still unclear as to whether the
separation between the molecular and metallic regions is sharp or
not. The planets are shown to scale, with their observed oblateness
and obliquity.}
\label{fig:jupsat}
\end{center}
\end{figure}

Figure~\ref{fig:rho_z} compares different pressure-temperature
profiles for Jupiter and Saturn, using the various equations of state
described here. The figure is intended to provide an estimate of
the uncertainties on the various equations of state (hydrogen-helium,
heavy elements). It is important to notice at this point that Saturn's
interior lies mostly in a relatively well-known region of the
hydrogen-helium EOS, i.e. in which hydrogen is molecular, whereas a
significant fraction
of Jupiter's interior is at intermediate pressures (one to a few Mbar)
for which the EOS is most uncertain.

\subsection{Opacities}

Although most of the interiors of Jupiter and Saturn are convectively
unstable, some regions have been recognized to be potentially {\it
stable}. At temperatures of the order of 1500 to 2000 K, a minimum of
the Rosseland mean opacity (see, e.g., Clayton 1968 for a definition) is
responsible for the likely presence of a
radiative region in Jupiter, and perhaps Saturn (Guillot et
al. 1994a). Much closer to the visible part of the atmospheres of
these planets, the underabundance of some compounds, in particular
water, may yield stable regions as well (see Guillot et al. 1994a;
Fig.5). Those may however not 
cover the entire surface of the planet at a given level. Instead, 2-
and 3-dimensional effects are to be expected due to meteorological
phenomena.

The presence of radiative regions has major consequences on both
static and evolution models of the giant planets: it yields
significantly cooler interiors (Guillot et al. 1994b), and quickens
their evolution (Guillot et al. 1995). It is therefore of prime
importance to calculate accurate radiative opacities. In this work, I
use a simplified opacity calculation, using the
method described in Guillot et al. (1994a), and updated absorption
data for water, methane, H$_2$S (see Marley et al. 1996 and references
therein). The results of these preliminary calculations are presented
in Table~\ref{tab:opa}, for a 3 times solar mixture. They are however
much improved compared to previous calculations (Guillot et
al. 1994a): as expected, including hot bands of water
increased the opacity by as much as a factor 3 in some places, thereby
significantly diminishing the extent and consequences of Jupiter's and
Saturn's inner radiative zones. In the case of Saturn, the present
models show no radiative zone, or a very limited one. The detailed
study of the new absorption data on the presence and magnitude of the
radiative zone will however have to await a more detailed calculation
of an extended Rosseland opacity table (Freedman, personal
communication 1999).

\begin{table}[htpb]
\begin{center}
\caption{Rosseland opacities (Log$\kappa_{\rm R}$, in g\,cm$^{-2}$)}
\label{tab:opa}
\begin{tabular}{crrrrr}\hline
&\multicolumn{5}{c}{\bf LogP (bar)} \\
{\bf LogT (K)} & \multic{0.0} & \multic{1.0} & \multic{2.0} &
\multic{3.0} & \multic{4.0} \\ \hline
2.2 & -1.68 &  -0.88 &   0.08 &   1.05 &  2.04 \\
2.4 & -1.75 &  -1.10 &  -0.29 &   0.50 &  1.32 \\
2.6 & -1.19 &  -0.58 &  -0.24 &   0.13 &  0.79 \\
2.8 & -1.08 &  -0.57 &  -0.30 &   0.09 &  0.81 \\
3.0 & -1.26 &  -0.89 &  -0.50 &   0.06 &  0.77 \\
3.2 & -2.17 &  -2.01 &  -1.68 &  -1.36 & -0.85 \\
3.4 & -1.93 &  -1.83 &  -1.66 &  -1.34 & -0.53 \\
3.6 & -1.02 &  -0.87 &  -0.72 &  -0.55 &  0.06 \\ \hline
\end{tabular}
\end{center}
\end{table}

\section{The three-layer structure assumption}

Jupiter and Saturn are thought to be fluid and mostly convective
(e.g., Hubbard 1968, Stevenson \& Salpeter 1977, Guillot et al. 1994a),
a consequence of
their significant intrinsic heat flux (Hanel et al. 1981, 1983). 
It is natural to infer that their interiors should be
homogeneously mixed, except where physical barriers prevent this
mixing. It is thus commonly accepted that these planets should consist
of a central core, a surrounding envelope in which hydrogen is
metallic, and an overlaying molecular envelope which would have a
chemical composition similar to what can be observed from the Earth
(or rather, from deep probes). The division between each region is, in
reality, rather vague. Figure~\ref{fig:jupsat} sketches the present
perception of 
the interiors of Jupiter and Saturn. The following
discussion aims at a careful examination of the
three-layer assumption, and its limits.

\subsection{The molecular envelope}

The molecular envelope is the region that extends from the
``surface'' (since there is no gas/liquid or gas/solid phase
transition, this usually refers to the visible troposphere, e.g., the 1
bar level), down to the level where hydrogen, instead of being molecular,
becomes metallic. It is assumed to be quasi-homogeneous, and
quasi-adiabatic, as discussed below.

With a pressure varying by more than six orders of magnitude from top
to bottom, the molecular region is necessarily complex, thus casting
some doubts on the assumption that it should be homogeneous. This is
best exemplified by Jupiter's visible atmosphere. Due to
condensation processes, the abundance of condensible species (NH$_3$,
H$_2$O) varies both vertically and horizontally (Niemann et al. 1998;
Roos-Serote et al. 1998). Similarly, complex chemical reactions occur
(Fegley \& Lodders 1994). However, since we are only interested in
those chemical species 
that contribute significantly to the total fluid density (i.e., H$_2$O,
essentially) these variations are expected to be confined to the upper
part of the envelope and to have only small consequences on the
calculated gravitational moments. They were however included in the
present work, the effect being mimicked by using the saturation vapor
pressure for water. The presence of refractory materials condensing at
deeper levels was neglected, as their abundances are significantly
smaller.

Condensation also tends to modify the temperature gradient,
due to the latent heat released by condensing particles. Although the
Galileo probe {\it in situ} measurements indicate that Jupiter's
temperature profile is consistent with a dry adiabat, it is not clear
that this is still valid out of the hot spot region (e.g., Showman \&
Ingersoll 1998). In fact, radio occultation data for Uranus and
Neptune show that significant departures from a dry adiabat
(generally steeper temperature profiles) occur in the region of methane
condensation, and can amount to variations of the order of 5-10\,K
(Lindal 1992). This is interpreted as a consequence of
the inhibiting action of molecular weight layering on convection
(Guillot 1995). Latent heat release and
molecular weight layering acting in opposite directions, it is not
possible to tell whether the interiors of Jupiter and Saturn might be
warmer or cooler than calculated assuming dry adiabatic profile. 
However, since they are expected to contain a much lower proportion of
heavy elements than Uranus and Neptune, the changes on the temperature
profile should not be as pronounced.
Finally, it is noteworthy that upper regions in which the abundance
of water is small, as was measured by the Galileo probe (Niemann et
al. 1998), might be convectively stable (Guillot et al. 1994a, Seiff
et al. 1998), which could also affect the temperature gradient. 
In summary, the temperatures in the 10-100 bar region and below appear
to be more uncertain than those measured at 1 bar, but I regard the
chosen 5 to 10\,K uncertainty on the 1\,bar temperature adequate as an
estimate of the variations due to the temperature.

Opacity calculations predict the presence of a radiative
region in Jupiter, at pressures between about 1.5 and 6 kbar, and
temperatures between 1450 and 1900 K, and a tiny one, if any in
Saturn. The presence of such a radiative region supposedly only
affects the temperature profile, not the chemical composition. This is
because any enrichment in heavy elements near the surface would be
rapidly transported by downwelling ``salt finger''-type
plumes. Furthermore, evolution calculations (Guillot et al. 1995) show
that this radiative zone, being essentially tied to the 1500-2000 K
temperature level, moves inward in time, which has the tendency to
smear any compositional gradient away. Finally, it is estimated that
meridional circulation and differential rotation should allow further
mixing of the elements in the radiative region. 

In summary, the molecular envelopes of Jupiter and Saturn are supposed
to be (dry) adiabatic, except where radiative opacities are low
enough, and homogeneous. This means that, except for species that
condense, the composition of the atmosphere is expected to be relevant
to that of the entire molecular envelope. In particular, this is true
of helium, whose measured mass mixing ratios in Jupiter and Saturn
are smaller than inferred in the protosolar nebula. Helium is therefore
believed to be hidden deeper, in the metallic region.

\subsection{The metallic envelope}

The distinction between the molecular and the metallic region is
difficult to make, except in the presence of the PPT which then would
have two physical effects: it would create
an effective entropy barrier which wouldn't be crossed by convection
(Stevenson \& Salpeter 1977). It would also induce a discontinuity of
the chemical composition, a consequence of the equality of the
chemical potentials of the two phases (see Landau \& Lifchitz
1984; Hubbard 1989). The molecular and metallic regions would
then be relatively
sharply separated, thus validating the present three-layer approach. 
The relative depletion of helium in the molecular regions of Jupiter
and Saturn could then also be due to this chemical abundance discontinuity
at the PPT, although this is generally not the preferred explanation
because helium is expected to be more soluble in molecular 
than in metallic hydrogen (Stevenson \& Salpeter 1977). 

In the absence of a PPT, the presence of a hydrogen/helium phase
{\it separation} becomes the only viable explanation for the low atmospheric
$Y$ in Jupiter and Saturn. Even if the PPT exists, a phase separation is
likely to occur, both in Jupiter and Saturn, although it is expected
that it is on larger scales inside Saturn because the planet is
colder, has less helium in its atmosphere, and because homogeneous
evolution models predict present day intrinsic heat flux much smaller
than observed. In that case, it is estimated that helium-rich droplets
would form, and grow rapidly, so that they would fall towards the
interior without being efficiently transported by convection
(Stevenson \& Salpeter 1977). Furthermore, this phase
separation is expected to lie close to the molecular/metallic
transition (e.g., Stevenson 1982). In this work, I choose to assume
that it occurs in a narrow region too, so that the helium-rich region
is essentially the metallic part of the envelope. Consequently, I
assume that, like the molecular region, the metallic one is homogeneous
and adiabatic. The temperature and pressure are 
continuous across the transition, but not necessarily the chemical
abundances, density and entropy.

In reality, hydrogen/helium separation may occur over a relatively
wide range of pressure. Figure~\ref{fig:jupsat} shows
that in the Mbar range, a relatively modest pressure variation from
the helium poor to the helium rich regions can involve a
non-negligible fraction of the planetary interior. It might even
concern most of the metallic region if the slope of the critical line
is as predicted by Pfaffenzeller et al. (1995) (see Guillot et
al. 1995). Furthermore, even in the case of a shallow inhomogeneous
region (with a gradient of helium concentration), a steep temperature
gradient, which would be treated as a discontinuity of the temperature
under the present assumptions, is to be expected. This jump should be
relatively small however, according to the estimates by Stevenson \&
Salpeter (1977). Moreover, the density of metallic hydrogen is only
weakly dependent on the temperature, and it can hence be neglected.

In summary, both major assumptions about the interior, i.e. that it is
adiabatic and homogeneous, 
could be wrong in a (significant) part of the metallic
region. Even in that somewhat least favorable case, the three-layer
models are not expected to be fully irrelevant because the
gravitational moments only probe extended regions of the interior
(e.g., Zharkov \& Trubitsyn 1974), and
without other information such as measurements of oscillations, only
constraints on averaged quantities can be sought from interior
models.

\subsection{The core}

As we shall see, it is possible (although improbable)
that Jupiter and Saturn have no cores. If present, the core has either been a
seed on which the giant planets have grown (Lissauer 1993), or it has
been formed early at the center of a very tenuous, non-convective
protoplanetary clump (Boss 1998). 
Its composition is unknown, and cannot be inferred from
the measurements of the gravitational moments (Guillot et
al. 1994b). I somewhat arbitrarily assume that it is formed of a
central mixture of rocks surrounded by ices, the ice to rock ratio
being variable.
I also assume that the transition between the  
core and the envelope is abrupt, although it might not be the case,
depending on what happened during the formation period (Stevenson
1985). 
If layering occurs in the central regions, an
intermediate inhomogeneous conductive region will link the core to the
metallic region. It is unclear whether, in the framework of our
three-layer model, the heavy elements present in such a region would
be mostly included in the metallic region, or as core mass.
In any case, the core masses given in this article are to be
interpreted as masses of heavy elements located at or near the
planetary center.

\section{Construction of models}

Interior models of the giant planets are calculated with the same set
of equations that govern the structure and evolution of stars, with
the exception that no thermonuclear reactions occur:
\begin{eqnarray}
\disp\dpar{P}{M}&=&-\frac{GM}{4\pi R^4}+{\omega^2\over 6\pi R}
+\frac{GM_{\rm tot}}{4\pi R^3_{\rm tot}R}{\varphi}_\omega,\label{eq:P}\\
\disp\dpar{T}{M}&=&\left(\dpar P M\right){T\over P} \nabla_{\!T} , \label{eq:T}\\
\disp\dpar{R}{M}&=&{1\over 4\pi R^2\rho} ,\label{eq:R}\\ 
\disp\dpar{L}{M}&=&-T\dpar S t,\label{eq:L}
\end{eqnarray}
where the variables are: $M$, mass; $R$, mean radius; $M_{\rm tot}$,
$R_{\rm tot}$: mass and radius at the ``surface''; $t$, time; $P$,
pressure; $T$, temperature; $\rho$, density; $L$, intrinsic luminosity; $S$,
specific entropy; $G$, gravitational constant; $\nabla_{\!T}=\partial\log
T/\partial\log P$, temperature gradient; $\omega$ is the angular velocity
and ${\varphi}_\omega$ is a slowly varying function of the radius due to the
high order centrifugal potential calculated within the theory of
figures (Zharkov \& Trubitsyn, 1978). In the case of static models
(i.e. trying to reproduce the giant planets as we can observe them
today rather than their entire evolution), Eq.(\ref{eq:L}) is not
calculated. Instead, the intrinsic luminosity is fixed to its observed
value, an approximation which is entirely justified, even in the
presence of a radiative window (see Guillot et al. 1994b, 1995). 

Solving this set of equations requires, apart from the usual boundary
conditions (nullity of the luminosity at the center, and observed
temperature at a given pressure level), the knowledge of:
\\
(i) {\it the equation of state} $\rho(P,T,\{x_i\})$, where
$x_i$ denotes the fractional abundance of constituent $i$. In this
work, I use the hydrogen-helium EOS from Saumon et al. (1995) in the
molecular and metallic parts of the envelope, with the assumption that
heavy elements can be embedded into an equivalent helium mass fraction
which then accounts for their presence. 
\\
(ii) {\it the temperature gradient} $\nabla_T$. It is determined by solving
the radiative/convective equilibrium using the mixing length theory
(e.g., Kippenhahn \& Weigert 1991). In most of the interior, it is
essentially equal to the adiabatic gradient obtained from the EOS:
$\nabla_{\rm ad}\equiv (\partial\ln T/\partial\ln P)_S$.
\\
(iii) {\it the chemical composition}. Within the adopted assumptions, the
composition of the envelope is 
determined by only two quantities, $Y_Z^{\rm mol}$ and $Y_Z^{\rm
met}$, the equivalent helium mass mixing ratio in the molecular and
metallic regions, respectively. The composition of the core, of mass
$M_{\rm core}$, is determined by the mass fraction of ices $f_{\rm
ice}$, the rest being rocks.

\begin{figure}[ht]
\begin{center}
\psfig{figure=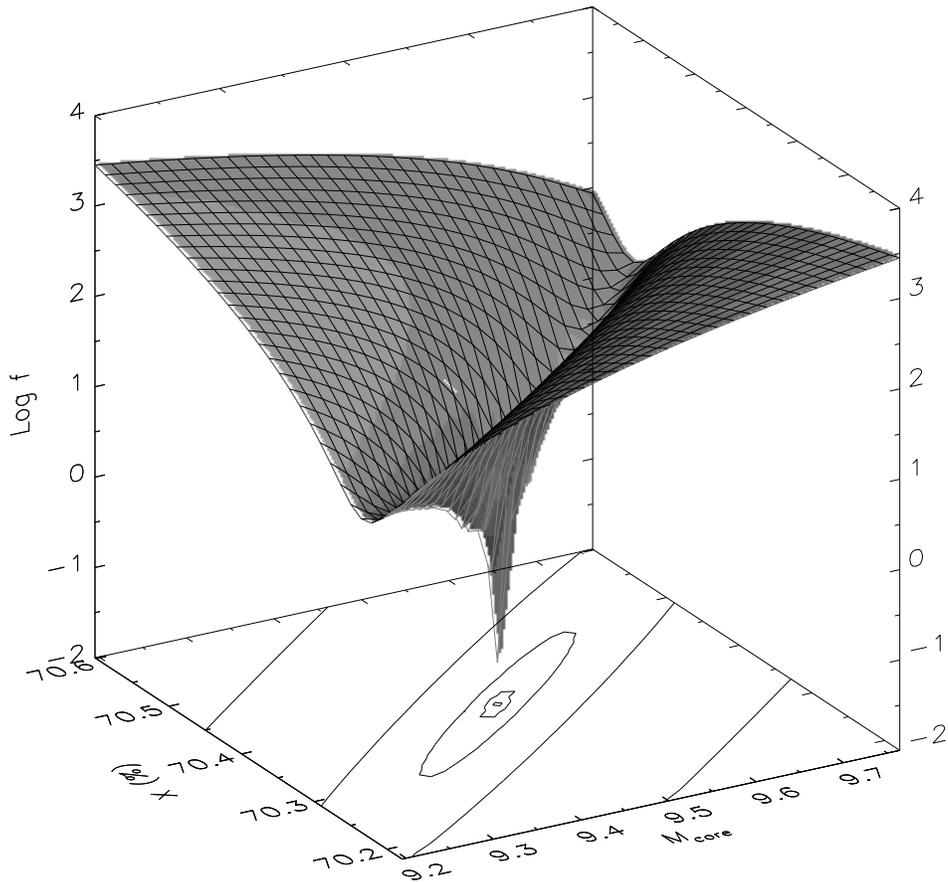,width=15cm}
\caption{Optimization of a model of Jupiter. The surface represents
the decimal logarithm of the minimization function $\chi^2$ (see
Eq.~\protect\ref{eq:optim}), as a function of the core mass $M_{\rm
core}$ and the hydrogen mass mixing ratio $X=1-Y_Z$ in the molecular
and metallic regions ($\Delta Y_Z=0$). Each point of the $\chi^2$
surface represents a converged interior model.}
\label{fig:opt_jup}
\end{center}
\end{figure}

\begin{figure}[ht]
\begin{center}
\psfig{figure=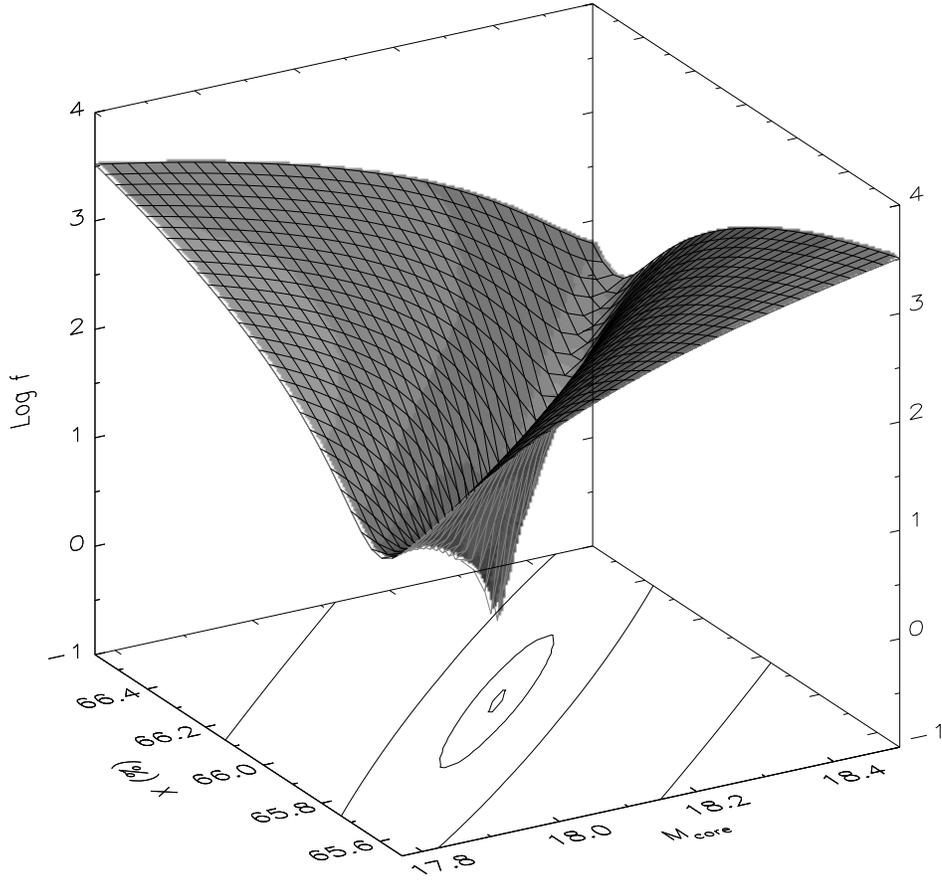,width=15cm}
\caption{Same as Fig.~\protect\ref{fig:opt_jup} for a model of
Saturn. Note that because one parameter is fixed ($\Delta Y_Z=0$), the
minimum found for $\chi^2(M_{\rm core},Y_Z^{\rm mol})$ is still, in
this case, $2.2\sigma$ away from the measured $J_4$.}
\label{fig:opt_sat}
\end{center}
\end{figure}

The total mass of the planet being fixed, the observational
constraints are the equatorial radius $R_{\rm eq}$ and gravitational
moments $J_2$, $J_4$ and $J_6$, measured with respective observational
uncertainties $\sigma_{R_{\rm eq}}$, $\sigma_{J_2}$, $\sigma_{J_4}$
and $\sigma_{J_6}$. In the framework of the three-layer
models, the adjustable parameters are $Y_Z^{\rm mol}$, $\Delta
Y_Z\equiv Y_Z^{\rm met}-Y_Z^{\rm mol}$ and $M_{\rm core}$. A way of
finding models matching the observational constraints is therefore to
minimize the following function:
\def\obs{{\rm obs}}
\def\rmc{{\rm calc}}
\begin{equation}
\chi^2(Y_Z^{\rm mol},\Delta Y_Z,M_{\rm core})={1\over 4}
\left[\left(\Delta R_{\rm eq} \over \sigma_{R_{\rm eq}}\right)^2+
\left(\Delta J_2 \over \sigma_{J_2}\right)^2+
\left(\Delta J_4 \over \sigma_{J_4}\right)^2+
\left(\Delta J_6 \over \sigma_{J_6}\right)^2\right],
\label{eq:optim}
\end{equation}
where $\Delta R_{\rm eq}$, $\Delta J_2$, $\Delta J_4$, $\Delta J_6$
are the differences between observed and theoretical $R_{\rm eq}$,
$J_2$, $J_4$ and $J_6$. 
For a given $\Delta X$, not too large in absolute value, it is generally
possible to find a well-defined, very sharp minimum of $\chi^2$
(Figs.~\ref{fig:opt_jup}, \ref{fig:opt_sat}). In fact, the minimum is
so sharp that the solution ($M_{\rm
core}$, $Y_Z^{\rm mol}$) can be thought of as being unique, i.e. leading
to an accurately defined value of $J_4$. The
choice of a different $\Delta Y_Z$ induces a different solution ($M_{\rm
core}$, $Y_Z^{\rm mol}$), and a different final value of $J_4$. The
non-uniqueness of solutions matching the observed gravitational fields
is therefore mostly due the uncertainty of $J_4$. So far, no
useful constraint can be derived from the values of $J_6$, owing to
their large observational uncertainties (see Table~\ref{tab:grav}). 

For each model solution, it is then possible to retrieve the mass
fraction of heavy elements in the molecular and metallic parts of the
envelope. This is done by using the observed helium mass fraction in
the molecular envelope $Y^{\rm mol}$ and by retrieving that in the
metallic envelope $Y^{\rm met}$ by using the fact that the total
helium to hydrogen ratio should be equal to its value in the
protosolar nebula. The value of Z at each level is then obtained by
the following relation:
\begin{equation}
Z=(Y_Z-Y){\rho_{\rm H}^{-1} - \rho_{\rm He}^{-1} \over \rho_{\rm
H}^{-1} - \rho_{\rm Z}^{-1}},
\label{eq:Z}
\end{equation}
where $\rho_{\rm H}$, $\rho_{\rm He}$ and $\rho_{\rm Z}$ are the
densities of pure hydrogen, pure helium and heavy elements,
respectively. As described by Guillot et al. (1997), Eq.~\ref{eq:Z}
does not ensure that $Z$ is constant, even in regions where $Y_Z$ and
$Y$ are. These departures having no physical meaning, $Z^{\rm mol}$
and $Z^{\rm met}$ are estimated by averaging Eq.~\ref{eq:Z} over the
molecular and metallic regions, respectively. 

Finally, in order to obtain meaningful constraints on the internal
composition of the giant planets, the following uncertainties have to
be taken into account:

{\it On the equations of state}: two EOSs from Saumon et al. (1995)
are used for the hydrogen-helium mixture. Because they are both
qualitatively and quantitatively very different from one another,
their use should provide a relatively good estimate of the
uncertainties that are to be expected on the behavior of these
elements at high pressures. 
On the other hand, the EOS for heavy elements is even more
uncertain because both their composition and high pressure behavior
are mostly unknown. The estimated uncertainties on $\rho_Z$ (that
enters Eq.~\ref{eq:Z}) are shown in Fig.~\ref{fig:rho_z} (see Guillot et al. 1997
for a more detailed description). 

{\it On the temperature profile}: the interior was either assumed to
be fully adiabatic, or calculated by consistently solving for the
radiative/convective equilibrium, thus allowing for a radiative region
to appear. 

{\it On the internal rotation}: Hubbard (1982) has shown that the
presence of differential rotation can lead to a significant change in
the calculated gravitational moments. This change is, in terms of
observational error, most important for the lower order
moments. In fact, the effect on $J_2$ leads only to a negligible
modification of the values of $Y_Z^{\rm mol}$, $\Delta Y_Z$ and
$M_{\rm core}$ that reproduce the observed $J_{2i}$. However, the
modification of $J_4$ by differential rotation is very significant, as
shown hereafter.

\section{Results}

\subsection{Sensitivity to changes of the density profiles}

As discussed previously, the density profile in the planetary interior
can vary, due to several causes (equation of state, temperature,
composition). However, only the density profile itself is constrained
by the gravitational field measurements. It is therefore useful
to estimate the effect of a change of the density profile on the
various parameters and results of the optimization procedure.

\begin{figure}[ht]
\begin{center}
\psfig{figure=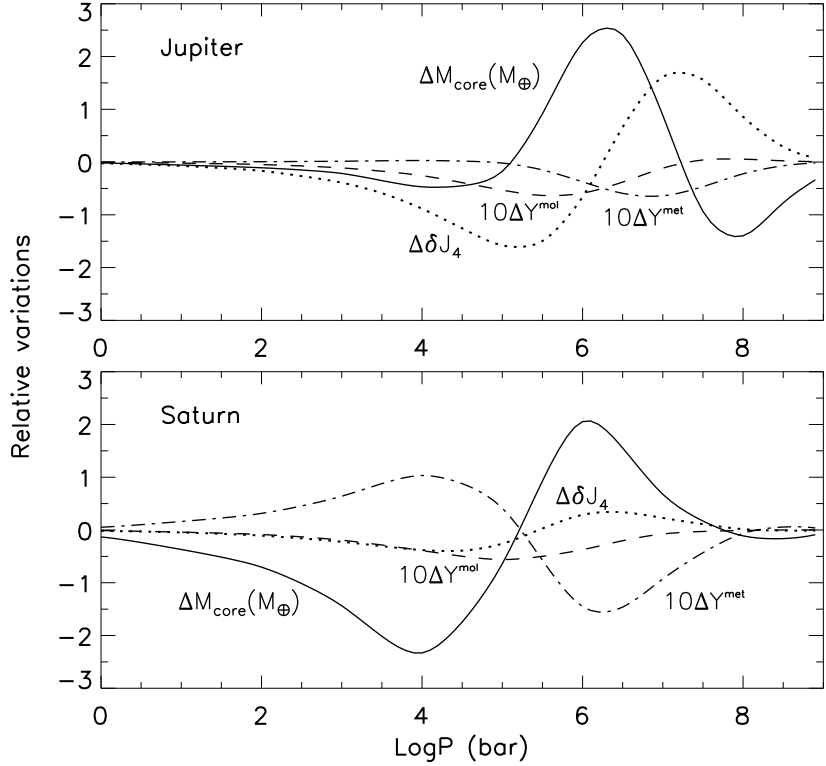,width=16cm}
\caption{Variations of $\delta J_4=[J_4-J_4({\rm observed})]/\sigma_{J_4}$, $M_{\rm
core}$ (in Earth masses), $Y_Z^{\rm mol}$ and $Y_Z^{\rm met}$  when
the density at a
given pressure $P_0$ (in abscissa) is increased by 5\%. The increase
is chosen to be a Gaussian in $\log(P/P_0)$ with a one-decade
half-width. The variations of $Y_Z^{\rm mol}$ and $Y_Z^{\rm met}$ have
been exaggerated by a factor 10.}
\label{fig:var_rho}
\end{center}
\end{figure}

In order to do so, two reference models are calculated for Jupiter and
Saturn. Comparison models are then computed, with a slightly
different EOS: instead of a density $\rho_0(P,T)$, I use $\rho(P,T)$,
which has been arbitrarily increased compared to $\rho_0$ by 5\% at
$P_0$, using the following relation:
\begin{equation}
\rho(P,T)=\rho_0(P,T)\left[1+0.05 e^{-\log^2(P/P_0)}\right],
\end{equation}
The comparison models are optimized by varying $Y_Z^{\rm mol}$ and
$M_{\rm core}$ so that $R_{\rm eq}$ and $J_{\rm 2}$ are always equal
to their observed value (of course, the total mass of the planet is
held constant!). The variations of $J_4$ that result are shown
by a dotted line on Fig.~\ref{fig:var_rho}. They can amount to up to
2$\sigma$ for Jupiter and 0.5$\sigma$ for Saturn (but the uncertainty on
the measured value of $J_4$ is larger). For example, a model of
Jupiter which at first fitted the observed gravitational moments
gets its $J_4$ shifted by 2$\sigma$ after a 5\% increase of the
density in the 10 Mbar region.

The additional constraint that $J_4$ should also remain fixed requires
an additional free parameter: $\Delta Y_Z$. A unique solution ($M_{\rm
core}$, $Y_Z^{\rm mol}$, $Y_Z^{\rm met}$) can then be sought. Its
changes as a function of $P_0$ are shown in Fig.~\ref{fig:var_rho}. In
both Jupiter and Saturn, up to 2\,$M_\oplus$ variations of the core masses
are observed in consequence to a localized 5\% density change. Note,
however, that a 5\% density change everywhere (i.e. independant of
$P$) would only scarcely affect the core mass, as can be seen by
integrating over $\log P$ the plain curve in
Fig.~\ref{fig:var_rho}. Jupiter and Saturn react differently to a
density change at pressure $P_0$: inside Jupiter, a density change at
pressures lower than 0.1\,Mbar involves only the outermost 5\% (in
radius) of the planet and is therefore barely seen. In Saturn, this
corresponds to more than 11\% of the planet. Furthermore, the
molecular/metallic transition is at 80\% of the planetary radius in
Jupiter, and about 50\% in Saturn. The consequence is that, in terms
of gravitational moments, Saturn's metallic region is essentially
indistinguishable from a large core. This is also seen from the large
variations of $Y_Z^{\rm met}$ with $P_0$ for this planet.

\subsection{Constraints from the gravity field}

\begin{figure}[ht]
\begin{center}
\vspace*{-1cm}
\centerline{\psfig{figure=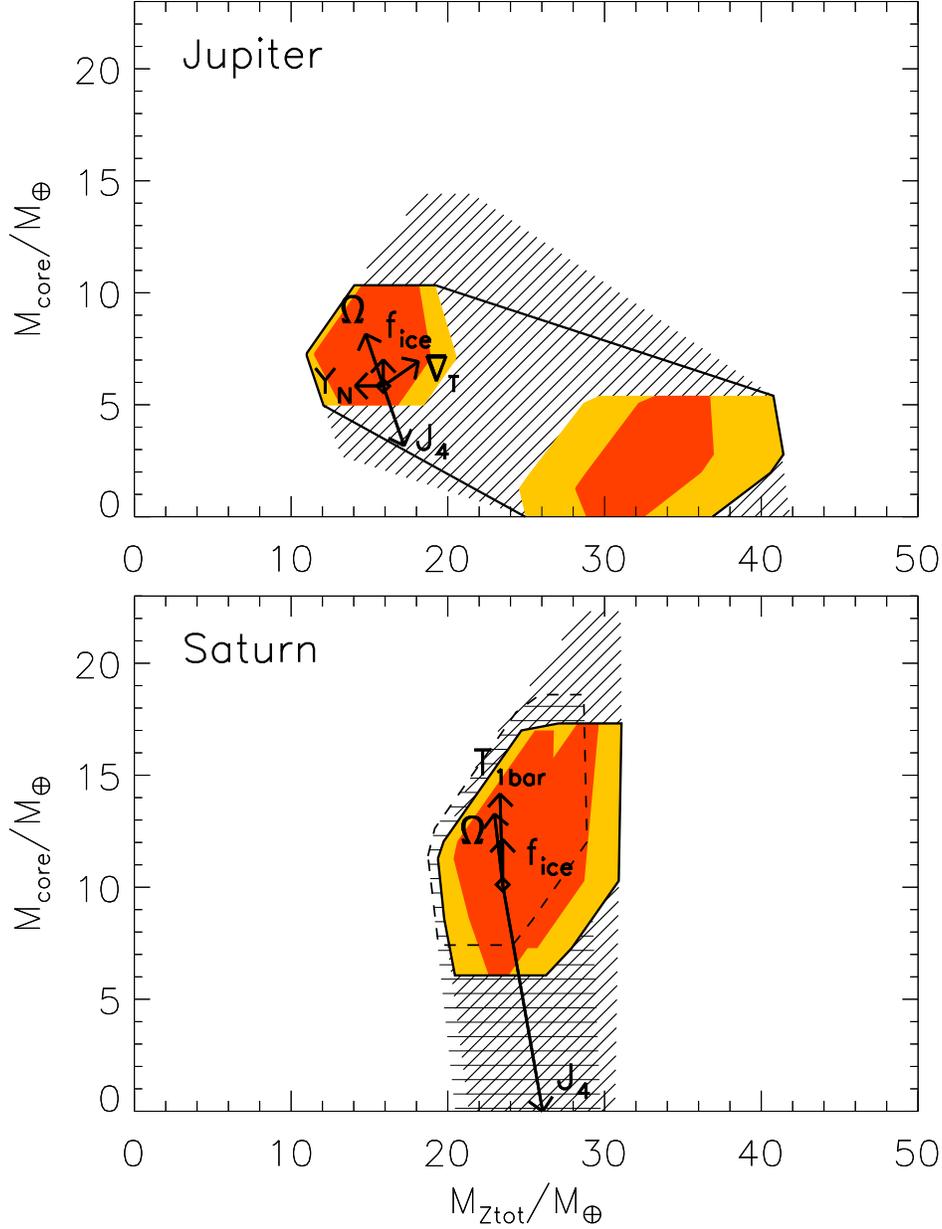,width=13cm}}
\vspace*{-1.5cm}
\caption{Constraints on Jupiter's (upper panel) and Saturn's (lower
panel) core masses ($M_{\rm core}$) and total masses of heavy elements
($M_{Z\rm tot}$), expressed in Earth masses ($M_\oplus$). The dashed
regions correspond to the constraints
obtained from static models only. The shaded regions are those which
also satisfy the
condition that the model ages should be close to the age of the Solar
System (see text). Within those, darker regions corresponds to
calculations that ignore uncertainties on the equation of state of
heavy elements. Models calculated with the PPT-EOS are to the left,
models using the $i$-EOS to the right. The region surrounded by the
thick line corresponds to the most plausible ($M_{\rm core}$, $M_{Z\rm
tot}$) region given all possible constraints and uncertainties. In the
case of Saturn, the horizontally-dashed region and the region
surrounded by dashed lines correspond to constraints using the Voyager
helium mixing ratios $Y=0.06\pm 0.05$, whereas other models assume
$Y=0.16\pm 0.05$. Arrows indicate the direction and magnitude of the
assumed uncertainties, if $J_4$ or $Y_{\rm proto}$ are increased by
$1\sigma$, rotation is assumed to be solid (``$\Omega$''), the core is
assumed to be composed of ices only (``$f_{\rm ice}$''), if Jupiter's
interior becomes fully adiabatic (``$\nabla_T$''), and if Saturn's
surface temperature is increased from 135 to 145 K (``$T_{\rm
1bar}$'').}
\label{fig:mcore}
\end{center}
\end{figure}

\begin{figure}[ht]
\begin{center}
\vspace*{-1cm}
\centerline{\psfig{figure=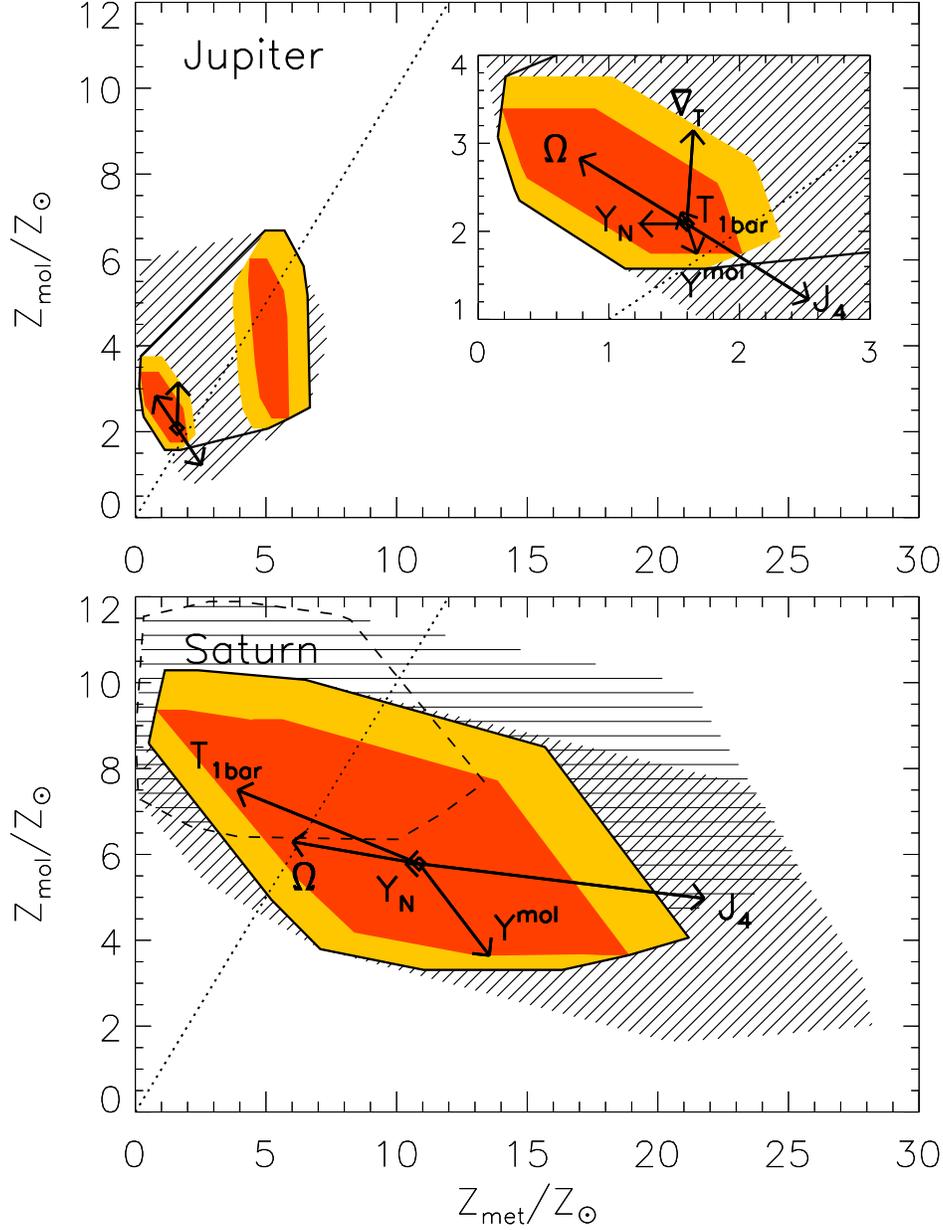,width=13cm}}
\vspace*{-1cm}
\caption{Same as Fig.~\protect\ref{fig:mcore} but for constraints on
the mass fractions of heavy elements in the molecular envelope
($Z_{\rm mol}$), and in the metallic envelope ($Z_{\rm met}$). The
$Z_{\rm mol}=Z_{\rm met}$ relations are shown by diagonal dotted
lines. Additional arrows not included in Fig.~\protect\ref{fig:mcore}
show the changes due to a 1-$\sigma$ increase in the surface helium
mass mixing ratio (``$Y^{\rm mol}$''), and in the case of Jupiter, of
an increase in surface temperature from 165 to 170K (``$T_{\rm
1bar}$''). The mixing ratios are in solar units ($Z_\odot=0.0192$; see
Anders \& Grevesse 1989). {\it Inset}: closeup view of the solutions
and assumed uncertainties for Jupiter models calculated with the
PPT-EOS.}
\label{fig:zmolmet}
\end{center}
\end{figure}

The constraints obtained solely from the static models are shown by
the dashed regions of Fig.~\ref{fig:mcore} and \ref{fig:zmolmet}. As
expected, the uncertainties on the hydrogen-helium EOS are much more
apparent in Jupiter than in Saturn, especially when one considers the
core masses and total masses of heavy elements obtained. In Jupiter,
the solutions obtained with the $i$-EOS and with the PPT-EOS do not
overlap. I however consider that all the solutions inbetween are
acceptable, as these two EOSs are characteristic of the uncertainties
of the high-pressure EOSs in general. In the case of Saturn, the
PPT-EOS yields generally smaller quantites of heavy elements, mostly 
in the metallic region, but the difference with the $i$-EOS is of the
order of 3\,$M_\oplus$ only. 

As shown in Fig.~\ref{fig:mcore}, the core masses obtained are
$0-14\,M_\oplus$ for Jupiter and $0-22\,M_\oplus$ for Saturn. The
total masses of heavy elements in these planets are $11-42\,M_\oplus$
and $19-31\,M_\oplus$, respectively. Unfortunately, it is not possible
to tell from this work which of Jupiter or Saturn contains the largest
amount of heavy elements, a fundamental question in the quest of their
origin. 

If the total amount of heavy elements is poorly constrained in
Jupiter, and rather well constrained in Saturn, the reverse is almost
true when considering the detailed enrichment in the molecular and
metallic envelopes, as seen on Fig.~\ref{fig:zmolmet}. As a result,
$Z^{\rm mol}$ is between 0.7 and 6.5
times solar, and $Z^{\rm met}$ is between 0 and 7 times solar in
Jupiter. The situation is slightly more complex in Saturn, because of
the uncertainty on the mass mixing ratio of helium in the atmosphere. 
The constraints resulting from the use of the Voyager helium mass
mixing ratios ($Y=0.06\pm 0.05$), shown as a horizontally-dashed area,
are, in solar units, $4.5<Z^{\rm mol}<13$ and $0<Z^{\rm
met}<25$. However, for reasons that will be
explained in the next section, a larger value of $Y^{\rm mol}$ is
probably more realistic. The ensemble of solutions for $Y^{\rm
mol}=0.16\pm 0.05$, $2\sigma$ away from the Voyager value, is
$2<Z^{\rm mol}<10.5$. This value is thus in better agreement
with the abundance of methane derived from spectroscopic data.
Unfortunately, no useful constraint on $Z_{\rm met}$ can be inferred
from these models. 

The results of Fig.~\ref{fig:mcore} agree fairly well with interior
models calculated 
within the last ten years: Hubbard \& Marley (1989) allowed the density
profile to vary in the molecular/metallic hydrogen transition region
and found that the more regular solutions had cores of the order of 8
to 12 M$_\oplus$ for Jupiter, and 9 to 20 M$_\oplus$ for Saturn. The
total masses of heavy elements that can be inferred from that work is
about 30 M$_\oplus$ and $20-30$ M$_\oplus$ for Jupiter and Saturn,
respectively. 
Zharkov \& Gudkova (1992), using five layers models, and a different,
fixed equation of state, found rock/ice core masses of about 5
M$_\oplus$ for Jupiter and 7 M$_\oplus$ for Saturn, these planets
containing about 50 M$_\oplus$ and 25 M$_\oplus$ of heavy elements,
respectively. Finally, Chabrier et al. (1992), with the same equations
of state as used by Guillot et al. (1994) and Guillot, Gautier \&
Hubbard (1997), found core masses of 4 to 8 M$_\oplus$ and 1 to 20
M$_\oplus$, and total masses of heavy elements of 10 to 16 M$_\oplus$
and 24 to 30 M$_\oplus$ for Jupiter and Saturn, respectively. 

Generally larger core masses (10-30 M$_\oplus$) were found in previous
calculations (see Stevenson 1982 for a review), but the largest core
masses also yielded helium mass fractions well below
the protosolar value that were therefore unrealistic. The main reason for
the discrepancy with today's values is however that the calculation
of core masses is, especially in the case of Jupiter, very sensitive
to changes in the equation of state, and that the core masses have
decreased as the equation of state of hydrogen has improved.

\begin{table}[hbt]
\begin{center}
\caption{Constraints on $\Delta Y_Z$ from static interior models}
\label{tab:deltayz}
\begin{tabular}{l*{2}{r@{ to }l}}\hline
 & \multicolumn{2}{c}{$i$-EOS} & \multicolumn{2}{c}{PPT-EOS} \\ \hline
Jupiter adiabatic 165K & 0.013&0.060 & $-0.051$&0.039 \\
\hphantom{Jupiter adiabatic} 170K & 0.003&0.060 & $-0.061$&0.029 \\
\hphantom{Jupiter} radiative 165K & 0.029&0.078 & $-0.030$&0.061 \\
\hphantom{Jupiter radiative} 170K & 0.003&0.078 & $-0.033$&0.057 \\
Saturn adiabatic 135K& 0.29&0.55 & 0.21&0.57 \\
\hphantom{Saturn adiabatic} 145K & 0.09&0.46 & 0.003&0.46 \\ \hline
\end{tabular}
\end{center}
\end{table}

An important parameter derived from these static models matching the
observed gravitational moments is the discontinuity of the equivalent
mass mixing ratio of helium $\Delta Y_Z$, as it can be linked to
results obtained from evolution models. The constraints interior
models provide to this parameter are given in Table~\ref{tab:deltayz}.

Finally, the relative importance of the various sources of
uncertainties included in this work on the solutions are indicated by
arrows in Figs.~\ref{fig:mcore} and \ref{fig:zmolmet}. One of the most
significant sources of uncertainty, apart from that relative to the
various equations of state, is that on the gravitational moment
$J_4$. A 1$\sigma$ change of this variable would amount to variations
of $M_{\rm core}$ by $\sim 3\rm\, M_\oplus$ in Jupiter, and by $\sim
10\rm\, M_\oplus$ in Saturn, and to large variations of of $Z^{\rm met}$
and $Z^{\rm mol}$. A much better determination of that
parameter by future missions (including, most importantly, the
Cassini-Huygens mission at Saturn) will therefore help to considerably
reduce the uncertainties on the model solutions. Unfortunately,
the presently unknown interior rotation field will set a limit on the
improvements that can be expected from such a determination, as shown
by the arrows labeled $\Omega$ in Figs.~\ref{fig:mcore} and
\ref{fig:zmolmet}. A better
determination of the opacities will also be a task to work on in
order to gain precision in the models.

In the case of Saturn, significant improvements are to be expected from
measurements by the Cassini-Huygens mission since 
the uncertainties associated with the equations of state are
relatively small for that planet. For example, a carefully
determined 1 bar temperature
will help to improve constraints on the core mass, $Z^{\rm mol}$
and $Z^{\rm met}$. A higher surface temperature (145K instead of 135K)
would thus yield a $\sim 4\rm\,M_\oplus$ larger $M_{\rm core}$,
values of $Z^{\rm mol}$ larger by 1.8 solar values, and of $Z^{\rm
met}$ smaller by 7 solar values. Furthermore, a better defined 
surface helium mass mixing ratio $Y^{\rm mol}$ is highly desirable, as
a 1$\sigma$ variation (with the current observational uncertainty)
yields an uncertainty on $Z^{\rm mol}$ which amounts to $\sim 2$ times
the solar value.

\subsection{Consistency with evolution models}

\begin{table}[hbt]
\begin{center}
\caption{Ages derived from homogeneous evolution models (in Gyr)}
\label{tab:evol}
\begin{tabular}{lcc}\hline
 & $i$-EOS & PPT-EOS \\ \hline
Jupiter adiabatic & 4.7 & 5.1 \\
\hphantom{Jupiter} radiative & 3.6 & 4.5 \\
Saturn adiabatic & 2.0 & 2.7 \\
\hphantom{Saturn} radiative & 2.0 & 2.2 \\ \hline
\end{tabular}
\end{center}
\end{table}

In this section, better constraints are sought from evolution
models, which are described in detail in the companion paper by
Hubbard et al. (1999). 
Model ages assuming {\it homogeneous\/} evolutions (see Guillot et
al. 1995), and corrected for the variable solar luminosity are given in 
Table~\ref{tab:evol}. Compared to the age of the Solar System, 4.55
Gyr, Jupiter and Saturn formed on a relatively short time scale, at
most 0.02 Gyr (e.g., Pollack et al. 1996). Any difference between the
ages in Table~\ref{tab:evol} and that of the Solar System has
therefore to be explained by either uncertainties inherent to the
evolution calculations (probably less than 0.3 Gyr), or by physical
phenomena, among which differentiation
appears to be the most likely. The time delay due to helium
differentiation can be estimated by the following relation (see
Hubbard et al. 1999):
\begin{equation}
\Delta t \simeq \left\{\begin{array}{ll} 
	9.1\, \Delta Y \qquad \mbox{for Jupiter,}\\
	8.3\, \Delta Y \qquad \mbox{for Saturn,}
	\end{array}\right.
\label{eq:deltat}
\end{equation}
where $\Delta t$ is in Gyr, and $\Delta Y$ could represent the
differentiation of either helium or any other elements (e.g., water, if
it is abundant enough). One can see that a negative $\Delta Y$ would
yield a faster
evolution, as heavy elements are then transported upward against
gravity, thereby consuming some of the internal luminosity. However,
this situation is extremely unlikely, as this would mean that helium
differentiation would be more than balanced by that of other elements,
which {\it a priori} are less abundant. I therefore assume that
$\Delta Y$ is positive. Some of the models of
Table~\ref{tab:evol}, in particular the adiabatic PPT-EOS Jupiter
evolution models, are thus incompatible with the age of the Solar System. 

Eq.~\ref{eq:deltat} assumes that $\Delta Y$ is a linearly increasing
function of time. This assumption fails in two cases: (i) if the
planet was initially formed with this abundance discontinuity or (ii)
if most of the evolution from homogeneity to the present $\Delta Y$
occurred early, as might happen with a compositional discontinuity
caused by the PPT. Eq.~\ref{eq:deltat} might hence overestimate
somewhat the time delay $\Delta t$, in which case larger $\Delta Y$
(or equivalently $\Delta Y_Z$) would then be possible. I choose
however, not to account for this possibility. The age constraints
$\Delta Y_Z$ are derived directly from Eq.~\ref{eq:deltat}, the
difference between the ages in Table~\ref{tab:evol} and that of 
the Solar System, assuming an uncertainty of $\pm 0.3$ Gyr for
Jupiter and $\pm 0.5$ Gyr for Saturn.

\begin{table}[hbt]
\begin{center}
\caption{Constraints on $\Delta Y_Z$ from evolution models}
\label{tab:ev_dyz}
\begin{tabular}{l*{2}{r@{ to }l}}\hline
 & \multicolumn{2}{c}{$i$-EOS} & \multicolumn{2}{c}{PPT-EOS} \\ \hline
Jupiter adiabatic  & $-0.05^\star$&0.01 & $-0.10^\star$&$-0.03^\star$ \\
\hphantom{Jupiter} radiative & 0.06&0.13 & $-0.03^\star$&0.03 \\
Saturn adiabatic & 0.24&0.36 & 0.15&0.28 \\
\hphantom{Saturn} radiative & 0.24&0.36 & 0.21&0.34 \\ \hline
\end{tabular}
\noindent
\parbox{9.5cm}{$^\star$: Negative values of $\Delta Y_Z$ are
unlikely (see text).}
\end{center}
\end{table}

The constraints thus obtained are summarized in
Table~\ref{tab:ev_dyz}. They can be compared
to the previous ones (Table~\ref{tab:deltayz}), the intersection of the
two providing a tighter global constraint. It is obvious, in the case
of Saturn, that significant helium differentiation is required
to provide the planet enough energy so as to slow its cooling by about
2 Gyr or more. However, large $\Delta Y_Z$ such as those predicted in
the case of models with no cores would provide too much energy, unless
most of the differentiation occurred early in the evolution, as
discussed below. 

An interesting side result can be derived with the additional
assumption that negligible differentiation of heavy elements occurs
between the molecular and the metallic regions, except any
compositional discontinuity that would occur during the formation
epoch, and would thus not affect Eq.~\ref{eq:deltat}. This assumption
is realistic, as helium is, due to its large abundance, the element
which is the most likely to separate from hydrogen (see, e.g., the
critical temperatures and abundances derived by Brami et al. 1979, in
the fully ionized regime). It is however not proven, and therefore
needs to be confirmed by experiments, or detailed theoretical calculations. 

With that assumption, the external helium abundance $Y^{\rm mol}$ can
then be derived. I write the constraint on the conservation of the
total quantity of helium:
\begin{equation}
Y^{\rm mol}=Y_{\rm proto}-{m_{\rm t}-m_{\rm c}\over 1 -m_{\rm c}} \Delta
Y,
\label{eq:ymol}
\end{equation}
in which $Y_{\rm proto}$ is the protosolar helium mass mixing ratio,
$m_{\rm c}$ the adimensional core mass and $m_{\rm t}$ the mass at the
molecular/metallic transition. I choose $m_{\rm t}\sim 0.85$,
$m_{\rm c}\sim 0$ for Jupiter and $m_{\rm t}\sim 0.5$, $m_{\rm c}\sim
0.1$ for Saturn. Table~\ref{tab:ymol} is then derived from
Eq.~\ref{eq:ymol} and the constraints from Tables~\ref{tab:deltayz}
and \ref{tab:ev_dyz}. It shows that {\it radiative} models of Jupiter
globally agree with the observed helium mass mixing ratio, $Y=0.238$,
although this probably requires an EOS intermediate between the
$i$-EOS and the PPT-EOS. On the other hand, they would not have agreed
(or only marginally) with the previous value $Y=0.18\pm 0.04$. Similarly,
the values derived for
Saturn, between 0.11 and 0.21, disagree with the Voyager value
$Y=0.06\pm 0.05$. If helium separation occurs deeper in the planet,
the problem becomes even more acute: for example, with $m_t=0.3$, I
derive $Y^{\rm mol}=0.19$ to 0.25. Similarly, if any other element
separates from hydrogen as well, the additional energy would require
still larger values of $Y^{\rm mol}$. 
A higher value of the surface
(molecular) helium mass mixing ratio is therefore required, a problem
that will certainly be addressed by the Cassini-Huygens mission. 

\begin{table}[hbt]
\begin{center}
\caption{Derivation of the values of $Y^{\rm mol}$}
\label{tab:ymol}
\begin{tabular}{l*{2}{r@{ to }l}}\hline
 & \multicolumn{2}{c}{$i$-EOS} & \multicolumn{2}{c}{PPT-EOS} \\ \hline
Jupiter adiabatic  & 0.26&0.28 & \multicolumn{2}{c}{---} \\
\hphantom{Jupiter} radiative & 0.20&0.23 & 0.24&0.28 \\
Saturn adiabatic & 0.11&0.17 & 0.14&0.21 \\
\hphantom{Saturn} radiative & 0.11&0.17 & 0.12&0.18 \\ \hline
\end{tabular}
\end{center}
\end{table}

In Figs.~\ref{fig:mcore} and \ref{fig:zmolmet}, the constraints
derived from evolution models are represented by shaded regions. They
are clearly smaller than those derived from static models only. In the
case of Jupiter, two separated regions represent the solutions for the
two EOSs used here. Clearly, a better-defined EOS is needed. The core
mass is thus constrained to lie between 0 and 10 Earth masses, but the
constraints on the total mass of heavy elements are not improved. The
new constraints also indicate that Jupiter's molecular envelope is
enriched in heavy elements by a factor 1.5 to 6.5 compared to the
solar value, and its metallic envelope by at most a factor $\sim 6.5$
(Fig.~\ref{fig:zmolmet}).

On the contrary, significantly tighter constraints are obtained from evolution
models in the case of Saturn. A lower core mass limit of
6\,M$_\oplus$ can be inferred. Models with smaller cores require 
abundance discontinuities $\Delta Y_Z$ that would yield ages larger
than that of the solar system. This constraint is however less robust
than that derived from the gravitational moments only, as it could be
invalidated by an early differentiation that would not obey
Eq.~\ref{eq:deltat}. This is however unlikely, as it is difficult to
understand why such an early differentiation would not have occurred in
Jupiter as well. The upper core mass limit is 16 M$_\oplus$. Saturn's
core mass therefore appears to be larger than that of Jupiter, but
this could be an artifact due, first to the rather large error bars,
and second to the fact that helium differentiation may yield an
extended inhomogeneous region in Saturn's metallic region.

\begin{figure}[ht]
\begin{center}
\vspace*{-6cm}
\centerline{\psfig{figure=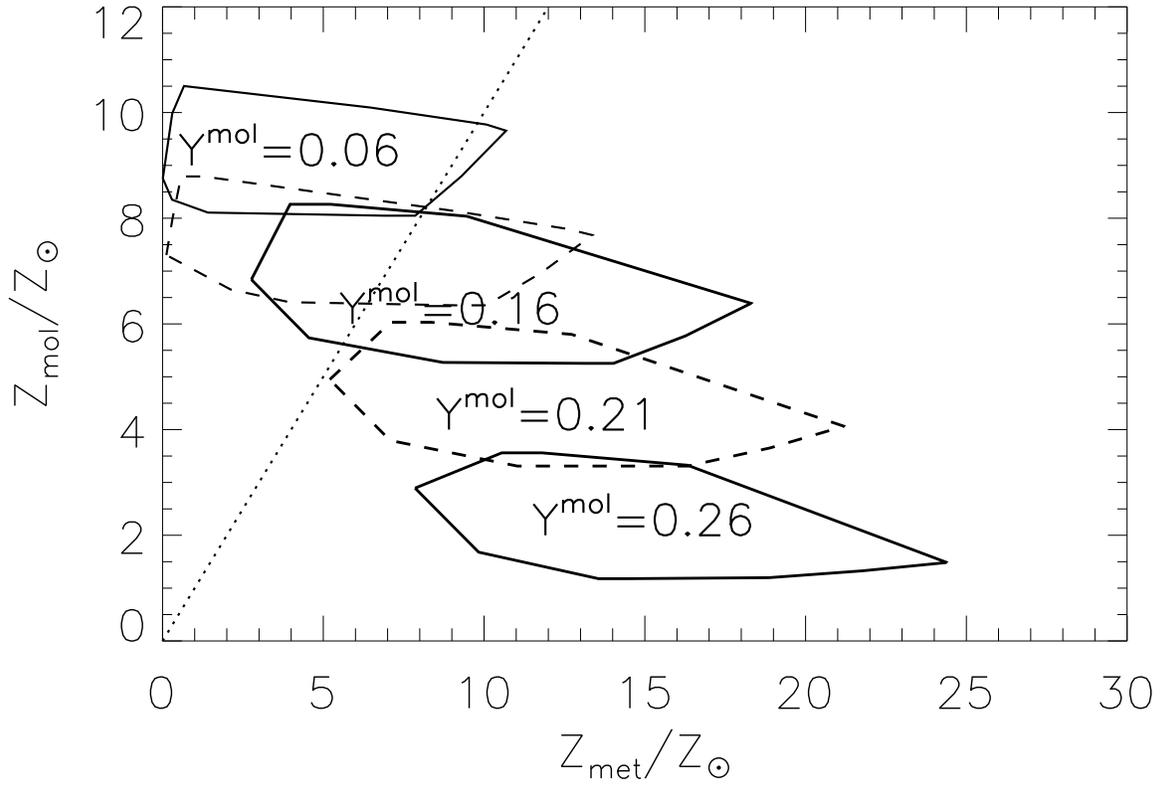,width=16cm}}
\caption{Same as Fig.~\protect\ref{fig:zmolmet} but for Saturn only.
The ensemble of solutions shown represent models matching gravitational
{\it and} age constraints for given surface helium mass mixing ratios, from
$Y^{\rm mol}=0.06$ to 0.26, as indicated (except for $Y^{\rm
mol}=0.11$, between $Y^{\rm mol}=0.06$ and 0.16). 
Observations of the abundances of chemical species in Saturn's
atmosphere tend to favor higher values of $Y^{\rm mol}$ than indicated
by Voyager ($Y=0.06\pm 0.05$).}
\label{fig:zsat}
\end{center}
\end{figure}

In Saturn, the case for a higher atmospheric helium mixing ratio than inferred
from Voyager is further strengthened by a derivation of $Z^{\rm mol}$,
shown in Fig.~\ref{fig:zsat}. A value
$Y^{\rm mol}$ less than 0.11 (the upper limit of the Voyager value
obtained by Conrath et al. 1984) yields enrichments in heavy elements
larger than 6.5 times the solar value, which is extremely
difficult to reconcile with the observed abundance of methane in that
atmosphere. The problem is of course complicated by the condensation
of many species to deeper levels, in particular water, and the
consequent impossibility to determine their abundances. However,
it seems difficult to imagine that the C/O ratio would be extremely
different from solar. On the contrary, higher values of $Y^{\rm mol}$
yield enrichments that agree with spectroscopic measurements of
Saturn's atmosphere. It is noteworthy that a high $Y^{\rm
mol}$ value necessitates that Saturn's metallic envelope be richer
in heavy elements than its molecular envelope. This presumably would
indicate the absence of a global mixing of the planetary interior
throughout its evolution, the inner regions, which formed early, having been
impacted by many more planetesimals in the early protoplanetary phases
than the final planet itself.

\subsection{The values of $J_4$ and $J_6$}

\begin{figure}[ht]
\begin{center}
\centerline{\psfig{figure=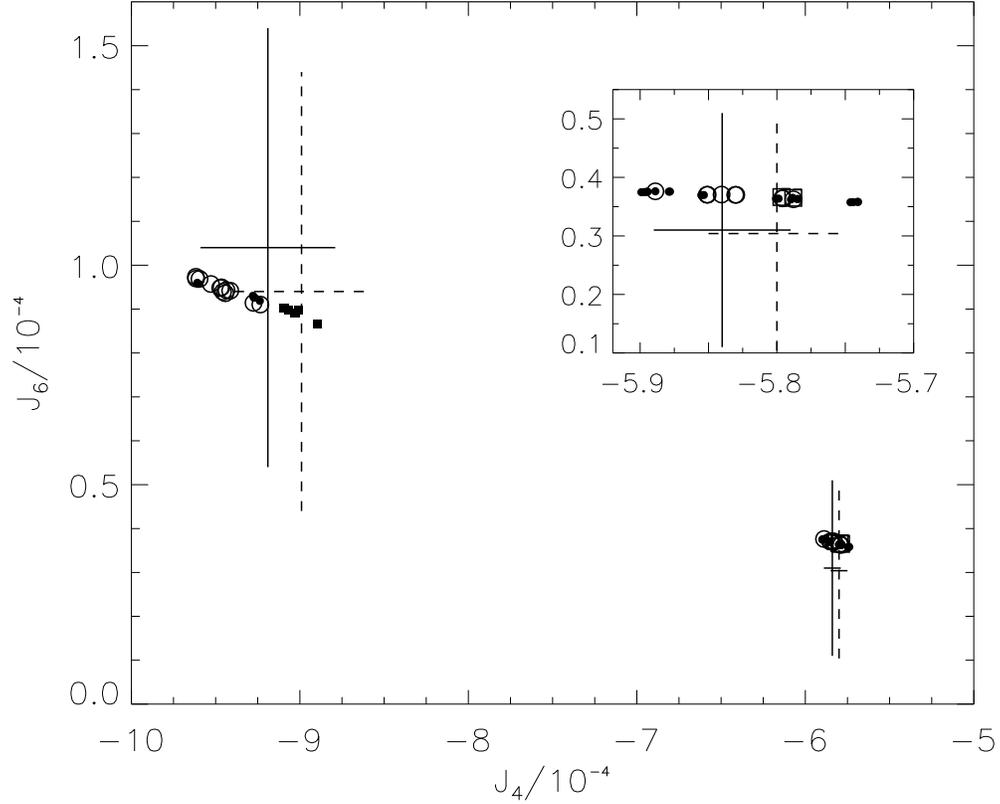,width=16cm,angle=90}}
\caption{Gravitational moments $J_4$ and $J_6$ of the optimized models of
Jupiter (right) and Saturn (left). Large open symbols represent models
that satisfy all constraints. Small filled symbols represent
models that match the observed gravitational field but do not predict
ages in agreement with that of the Solar System. Squares are models
with {\it no} core, all other models being represented by circles. The
plain crosses are the observational constraints on ($J_4$, $J_6$). The
dashed crosses show the constraints derived when differential rotation
is assumed to reach the deep interior. Inset is an enlargement of the
solutions obtained for Jupiter.}
\label{fig:j4j6}
\end{center}
\end{figure}

An interesting side result concerns the values of $J_4$ and $J_6$
obtained assuming solid rotation (only the {\it measured}
observational moments have been adjusted for differential rotation),
shown in Fig.~\ref{fig:j4j6}. The values of $J_4$ were constrained to
be within the observational error bars, but it can be seen that, both
in the case of Jupiter and Saturn, the models with the additional
evolution constraint (large symbols) are generally in better agreement
with a solid internal rotation than with a surface differential
rotation extending deep in the interior (Hubbard 1982). The
uncertainties on the measurements are however still too large to be
conclusive. Again, more accurate $J_4$ measurements may resolve
the issue of the presence of a core, as models with no core always
possess smaller $|J_4|$. 

From Jupiter to Saturn, a close proportionality relation links $J_6$
to $J_4$, almost independently of any assumed uncertainty (on the
equation of state, temperature profile, condensation of heavy
elements...etc.). This is not unexpected, as the uncertainties on the
density profile in the external layers are rather small, and these
profiles are continuous. The theoretical values for $J_6$ are between
0.35 and $0.38\times 10^{-4}$ for Jupiter, and between 0.9 and
$0.98\times 10^{-4}$ for Saturn. These values are thus defined with a
precision which is about 10 times larger than the current
observational uncertainties. Should measurements point to different
values of $J_6$, major changes in interior models would have to be
found, probably invoking significant departures from the three-layer
assumption, and/or very peculiar differential rotation patterns.
For example, an abstract by Bosh (1994) states that a more accurate
value of $J_6$ can be obtained for Saturn by constraints added by
measurements of the position of non-circular ringlets of the
planet. The thus inferred value (adjusted to the 1\,bar equatorial
radius given in Table~\ref{tab:grav}), $J_6=1.25\pm 0.06\times
10^{-4}$, is incompatible with all interior models presented
here. This has also been verified by Gudkova \& Zharkov (1997).

Measurements by the Cassini orbiter should provide accurate
measurements of $J_4$ and $J_6$, and therefore test interior
models. It also has been proposed that a Jupiter close orbiter would
allow the precise determination of the planet's gravitational moments
up to a relatively high order ($\sim 12$), and would thus constrain
the deep internal rotation (Hubbard 1999). Such a mission would also
provide an excellent test of interior models of Jupiter.

\subsection{The D/H ratios}

\def\dsurh{({\rm D/H})}

The measurement of the deuterium to hydrogen isotopic ratio in the
giant planets could be a powerful way to determine the abundance of ices in
their interiors, as this isotopic ratio $\dsurh_{\rm ices}$ is, in the
interstellar medium, higher 
in ices (HDO/H$_2$O) than in hydrogen (HD/H$_2$). The giant
planets are expected to contain slightly more deuterium than the protosolar
value $\dsurh_{\rm proto}$, in proportion to their enrichment in heavy
elements (Hubbard and MacFarlane 1980; L\'ecluse et al. 1996):
\begin{equation}
\dsurh = {M_{\rm H_2} \dsurh_{\rm proto} + {2\over 18} M_{\rm H_2O}
\dsurh_{\rm H_2O} \over M_{\rm H_2} + {2\over 18} M_{\rm H_2O}},
\end{equation}
where $M_{\rm H_2}$ and $M_{\rm H2O}$ are the masses of the hydrogen and water
reservoirs, respectively, that were allowed to exchange their deuterium, and
$\dsurh_{\rm H_2O}$ is the deuterium ratio in water whose
assumed value, 300\,ppm, is representative of that found in comets so far
(Bockel\'ee-Morvan et al. 1998). Other compounds may also have a small
contribution to the deuterium enrichment but are neglected. 

\begin{figure}[ht]
\begin{center}
\centerline{\psfig{figure=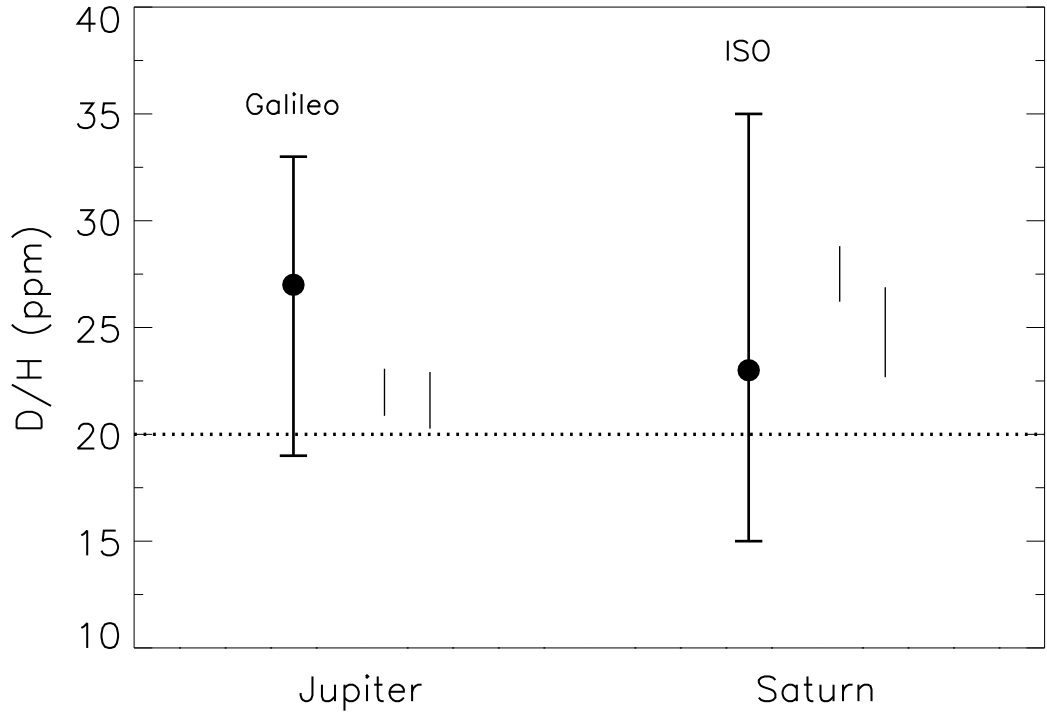,width=16cm}}
\caption{Observational and theoretical isotopic D/H ratio in Jupiter
and Saturn, in parts per million (ppm). The measurements from Galileo
in Jupiter and ISO in Saturn are shown with their error
bars. Theoretical estimates (thin lines) are calculated assuming
that ice carriers had a D/H similar to that observed in comets (300
ppm). The lines that are to
the left assume that the central core exchanged its deuterium with the
envelope, the lines to the right (which predict slightly smaller D/H
values) assume that such exchange didn't take place because of
convection inhibition at the core/envelope boundary. The assumed
protosolar D/H ratio used for the calculation (20 ppm) is shown as a dotted
line. The additional uncertainty on the protosolar value was not taken
into account. Its value is inferred from the $^3$He/$^4$He ratio in
the solar wind, and amounts to $21 \pm 5$\,ppm (Geiss \& Gloecker,
1998). Deuterium enrichment on Jupiter and Saturn is hence not
detectable yet.}
\label{fig:dsurh}
\end{center}
\end{figure}

Figure~\ref{fig:dsurh} compares the D/H ratios measured on Jupiter
(Mahaffy et al. 1998) and Saturn (Griffin et al., 1996) to the
theoretical values estimated from the interior models
matching both the gravitational and evolution constraints. 
Two theoretical calculations were performed for each planet, one for
which the total amount of ices was allowed to exchange its deuterium
with the hydrogen reservoir, another for which the core was kept
separated from the envelope (i.e. material in the core never mixed
with the hydrogen in the envelope, even during the formation of the
planets). 
The mass of water (and other deuterium-carriers) that contributed to
the heavy element content of the giant planets was supposed to lie
between 1/2 and 2/3 of the total mass of heavy elements now present in
these planets. 
As a result, even if Jupiter and Saturn have been enriched
in deuterium by ices with relatively high D/H ratios, this enrichment
is still too small to be detected, given the large observational error
bars. 

A better observational determination of the deuterium abundances in
Jupiter and Saturn is to be sought. If measured
with an accuracy of the order of 1ppm, a comparison
of the values obtained for the two planets would allow an
estimate of the deuterium content of the primordial ice carriers,
thus testing the hypothesis that ices that formed planetesimals
vaporized and had time to exchange their deuterium in the hot early
protosolar nebula (Drouart et al. 1999).

\section{Conclusions}

Constraints on the internal structure and composition of Jupiter and
Saturn have been derived within the framework of the three-layer
model. For the first time, constraints from both the planets'
gravitational fields and their evolution have been used. The
uncertainties of the result are still large, mostly because of our
limited knowledge of the behavior of hydrogen and of hydrogen/helium
mixtures at pressures greater that 1 Mbar. 

Yet interior models indicate that both Jupiter and Saturn are
enriched in heavy elements compared to the Sun (or, equivalently, the
protosolar nebula). How the planets acquired these additional heavy
elements is still a mystery. Unfortunately, a
comparison of core masses and amounts of heavy elements in Jupiter and
Saturn is still inconclusive. Most importantly, it is not yet possible
to decide which of Jupiter and Saturn possesses the largest absolute
quantity of elements other than hydrogen and helium. 

From this work, one cannot rule out either core accretion
or gaseous instability as valid formation mechanisms for
Jupiter. Uncertainties 
on the hydrogen equation of state
greatly affect the accuracy of the results obtained for that planet,
and both a low and a high heavy elements content (which, presumably would be
incompatible with a rapid formation of the planet, as its capture radius
would become small on time scales shorter than 1 Myr) are possible. 

In the case of Saturn, constraints from the interior and
evolution models, indicate that a significant fraction of the
heavy elements lie in the dense core and in the metallic envelope.
This is difficult to reconcile with the formation of this
planet by gaseous instability. 

Although, it will be difficult to accurately determine the structure of
the interiors of Jupiter and Saturn in the foreseeable future,
progress is to be expected from high pressure experiments on liquid
hydrogen (or deuterium) and the corresponding theoretical improvements
on the equation of state of that material, from better determinations
of Saturn's helium abundance in its atmosphere, from measurements of
the high order gravitational moments (beginning with $J_4$) in Jupiter
and Saturn, and from theoretical improvements on opacity data and
non-homogeneous evolution models including hydrogen/helium phase
separation.

\section*{Acknowledgments}

I wish to thank D. Gautier, W.B. Hubbard and D. Saumon for many
helpful discussions and e-mail exchanges, and T. Owen for carefully
reading this manuscript. 
This work was supported by the {\it Groupe de Recherche Structure
Interne des Etoiles et des Plan\`etes G\'eantes} and by the {\it
Programme National de Plan\'etologie}. It has been performed
using the computing facilities provided by the program ``Simulations
Interactives et Visualisation en Astronomie et M\'ecanique (SIVAM)''.

\section*{References}

\bib{Anders, E., Grevesse, N.} 1989. Abundances of the elements:
meteoritic and solar. {\it Geochim. Cosmochim. Acta} {\bf 53}, 197--214.

\bib{Bahcall, J.N., Pinsonneault, M.H.} 1995.
Solar models with helium and heavy elements diffusion.
{\it Rev. Mod. Phys.} {\bf 67}, 781--808.

\bib{Bockel\'ee-Morvan, D., Gautier D., Lis, D.C., Young, K., Keene,
J., Phillips, T., Owen, T., Crovisier, J., Goldsmith, P.F., Bergin,
E.A., Despois, D., Wootten, A.} 1998. Deuterated water in comet C/1996
B2 (Hyakutake) and its implications for the origin of comets. {\it
Icarus} {\bf 133}, 147--162.

\bib{Bosh, A.S.} 1994. Saturn's non-circular ringlets and gravitational
harmonics. {\it Ann. Geophys.} {\bf 12}, C676 [abstract].

\bib Boss, A. P. 1998. Evolution of the Solar Nebula IV - Giant
Gaseous Protoplanet formation. {\it \ApJ} {\bf 503}, 923--937.

\bib {Brami, B., Hansen, J-P., Joly, F.} 1979. Phase separation of highly
dissymmetric binary ionic mixtures. {\it Physica} {\bf 95A}, 505--525.

\bib Burrows, A., Marley, M.S., Hubbard, W.B., Lunine, J.I., Guillot, T.,
Saumon, D., Freedman, R., Sudarsky, D. \& Sharp, C. 1997. A nongray theory
of extrasolar giant planets and brown dwarfs. {\it Astrophys. J.} {\bf
491}, 856--875.

\bib{Busse, F.H.} 1976. A simple model of convection in the Jovian
atmosphere. {\it Icarus} {\bf 29}, 255-260.a

\bib{Cameron, A.G.W.} 1978. Physics of the primitive solar accretion
disk. {\it Moon Plan.} {\bf 18}, 5--40.

\bib {Campbell, J.K., Synnott, S.P.} 1985. Gravity field of the jovian system
from Pioneer and Voyager tracking data. {\it Astron. J.} {\bf 90},
364--372. 

\bib {Campbell, J.K., Anderson, J.D.} 1989. Gravity field of the saturnian
system from Pioneer and Voyager tracking data. {\it Astron. J.} {\bf
97}, 1485--1495.

\bib{Carlson, B.E., Lacis, A. A., Rossow, W. B.} 1992.
The abundance and distribution of water vapor in the Jovian troposphere 
as inferred from Voyager Iris observations. {\it \ApJ} {\bf 388}, 648-688.

\bib{Chabrier, G., Saumon, D., Hubbard, W.B., Lunine, J.I.} 1992. 
The molecular-metallic transition of hydrogen and the structure of Jupiter
and Saturn. {\it \ApJ} {\bf 391}, 817--826.

\bib{Clayton, D.D.} 1968. {\it Principles of stellar evolution and
nucleosynthesis}, Univ. Chicago Press, Chicago (Revised edition:
1983).

\bib{Collins, G.W., Da Silva, L.B., Celliers, P., Gold, D.M., Foord,
M.E., Wallace, R.J., Ng, A., Weber, S.V., Budil, K.S., Cauble, R.}
1998. Measurements of the equation of state of deuterium at the fluid
insulator-metal transition. {\it Science} {\bf 281}, 1179--1181.

\bib{Conrath, B.J., Gautier, D., Hanel, R., Lindal, G., Marten, A.}
 1984. The helium abundance of Saturn from Voyager measurements. 
{\it \ApJ} {\bf 282}, 807-815.

\bib{Conrath, B.J., Hanel, R.A.,Samuelson,  R.E.}
1989. Thermal structure and heat balance of the outer planets. In {\it
Origin and Evolution of Planetary and Satellite Atmospheres}
(eds. S.K. Atreya, J.B. Pollack, and M.S. Matthews), Univ. of
Arizona Press, Tucson, pp. 513--538. 

\bib{Da Silva, L.B., Celliers, P., Collins, G.W., Budil, K.S.,
Holmes, N.C., Barbee, III, T.W., Hammel, B.A., Kilkenny, J.D.,
Wallace, R.J., Ross, M., Cauble, R.} 1997. Absolute equation of state
measurements on shocked liquid deuterium up to 200 GPa (2 Mbar). {\it
Phys. Rev. Lett.} {\bf 78}, 483--486.

\bib{de\,Pater, I., Massie, S.T.} 1985. Models of the millimeter-centimeter 
spectra of the giant planets. {\it Icarus} {\bf 62}, 143--171.

\bib{Drouart, A., Dubrulle. B., Gautier, D., Robert,
F.} 1999. Structure and transport in the solar nebula from constraints
on deuterium enrichment and giant planets formation. {\it
Icarus}, in press. 

\bib{Fegley, B.Jr., Lodders, K.} 1994. Chemical models of the deep
atmospheres of Jupiter and Saturn. {\it Icarus} {\bf 110}, 117--154.

\bib{Folkner W.M., Woo, R., Nandi, S.} 1998. Ammonia abundance
in Jupiter's atmosphere derived from the attenuation of the Galileo
probe's radio signal. {\it \JGR} {\bf 103}, 22831--22846.

\bib{Gautier, D., B\'ezard, B., Marten, A., Baluteau, J.-P., Scott, N.,
Chedin, A., Kunde, V., Hanel, R.} 1982. The C/H ratio in Jupiter from the
Voyager infrared investigation. {\it \ApJ} {\bf 257}, 901--912.

\bib{Gautier D., Owen, T.} 1989. Composition of outer planet atmospheres.
In {\it Origin and Evolution of Planetary and Satellite Atmospheres}
(eds. S.K. Atreya, J.B. Pollack, and M.S. Matthews),
University of Arizona Press, Tucson, pp. 487--512. 

\bib{Geiss, J., Gloecker, G.} 1998. Abundances of deuterium and
helium-3 in the protosolar cloud. {\it Space Sci. Rev.} {\bf 84},
239--250.

\bib{Gierasch, P.J., Conrath, B.J.} 1993. Dynamics of the atmospheres
of the outer planets - Post-Voyager measurement objectives. {\it
J. Geophys. Res.} {\bf 98}, 5459-5469.

\bib{Griffin, M.J., Naylor, D.A.,
 Davis, G.R., Ade, P.A.R.,
 Oldham, P.G., Swinyard, B.M.,
 Gautier, D., Lellouch, E.,
 Orton, G.S., Encrenaz, T.,
 De Graauw, T., Furniss, H.,
 Smith, I., Armand, C.,
 Burgdorf, M., Di Giorgio, A.,
 Ewart, D., Gry, C., King, K.J.,
 Lim, T., Molinari, S., Price, M.,
 Sidher, S., Smith, A., Texier, D.,
 Trams, N., Unger, S.J., and
 Salama, A. 1996. 
First detection of the 56-{$\mu$}m rotational line of HD in Saturn's
atmosphere. {\it Astron. Astrophys.} {\bf 315}, L389-L392}

\bib{Gudkova, T.V., Zharkov, V.N.} 1997. Models of Jupiter and Saturn
with water-depleted atmospheres. {\it Solar Sys. Res.} {\bf 31},
99--107.

\bib{Guillot, T., Gautier, D., Chabrier, G., Mosser, B.} 1994a.
Are the giant planets fully convective? {\it Icarus}, {\bf 112},
337--353.

\bib{Guillot, T., Chabrier, G., Morel, P., Gautier, D.} 1994b.
Non-adiabatic models of Jupiter and Saturn. {\it Icarus}, {\bf 112},
354--367.

\bib{Guillot, T.} 1995. Condensation of methane, ammonia and water and the
inhibition of convection in giant planets. {\it Science} {\bf 269},
1697--1699.

\bib{Guillot, T., Chabrier, G., Gautier, D., Morel, P.} 1995. Radiative
transport and the evolution of Jupiter and Saturn. {\it
Astrophys. J.} {\bf 450}, 463--472.

\bib{Guillot, T., Gautier, D., Hubbard, W.B.} 1997. New constraints on the
composition of Jupiter from Galileo measurements and interior
models. {\it Icarus} {\bf 130}, 534--539.

\bib{Hanel, R.A., Conrath, B.J., Herath, L.W., Kunde, V.G., Pirraglia, J.A.}
1981. Albedo, internal heat, and energy balance of Jupiter:
Preliminary results of the Voyager infrared investigation. {\it J.
Geophys. Res.} {\bf 86}, 8705--8712.

\bib {Hanel, R.A., Conrath, B.J., Kunde, V.G., Pearl, J.C., Pirraglia, J.A.}
1983. Albedo, internal heat flux, and energy balance of Saturn. {\it
Icarus} {\bf 53}, 262--285.

\bib{Holmes, N.C., Ross, M., and Nellis, W.J.} 1995. Temperature
measurements and dissociation of shock-compressed liquid deuterium and
hydrogen. {\it Phys. Rev. B} {\bf 52}, 15835--15845.

\bib{Hubbard, W.B.} 1968. Thermal structure of Jupiter. {\it \ApJ}
{\bf 152}, 745--753. 

\bib{Hubbard, W.B., MacFarlane, J.J.} 1980. Theoretical predictions of
deuterium abundances in the jovian planets. {\it Icarus} {\bf 44},
676--682.

\bib{Hubbard, W.B.} 1982. Effects of differential rotation on the
gravitational figures of Jupiter and Saturn. {\it Icarus} {\bf 52},
509--515.

\bib{Hubbard, W.B.} 1989. Structure and composition of giant planets
interiors. In {\it Origin and Evolution of Planetary and Satellite 
Atmospheres}, S. K. Atreya, J. B. Pollack, and M. S. Matthews, eds.,
University of Arizona Press, Tucson, pp. 539--563.

\bib{Hubbard, W.B., Marley, M.S.} 1989. Optimized Jupiter, Saturn and
Uranus interior models. {\it Icarus} {\bf 78}, 102--118.

\bib{Hubbard, W.B., Guillot, T., Lunine, J.I.,
Burrows, A., Saumon, D., Marley, M.S., Freedman, R.S.} 1997.
Liquid metallic hydrogen and the structure of
brown dwarfs and giant planets,
{\it Physics of Plasmas}, {\bf 4}, 2011--2015.

\bib{Hubbard, W.B.} 1999. Gravitational signature of Jupiter's deep
zonal flows. {\it Icarus} {\bf 137}, 196--199.

\bib{Hubbard, W.B., Guillot, T., Marley, M.S., Burrows, A., Lunine,
J.I., Saumon, D.} 1999. Comparative evolution of Jupiter and
Saturn. Submitted to {\it Planet. Space Sci.}

\bib {Kippenhahn, R., Weigert, A.} 1991. {\it Stellar Structure 
and Evolution}, Springer-Verlag, Berlin.

\bib{Klepeis, J.E., Schafer, K.J., Barbee, T.W., III, Ross, M.} 1991.
Hydrogen-helium mixtures at megabar pressures: Implications for
Jupiter and Saturn. {\it Science} {\bf 254}, 986--989.

\bib{Landau, L., Lifchitz, E.} 1984. {\it Physique statistique}
(French translation), Editions Mir, Moscow. 

\bib{L\'ecluse, C., Robert, F., Gautier, D., Guiraud, M.}
1996. Deuterium enrichment in giant planets. {\it Plan. Space Sci.}
{\bf 44}, 1579--1592.

\bib{Lindal, G.F., Wood, G.E., Levy, G.S., Anderson, J.D., Sweetnam, D.N., Hotz,
H.B., Buckles, B.J., Holmes, D.P., Doms, P.E., Eshleman, V.R., Tyler, G.L.,
Croft, T.A.} 1981. The atmosphere of Jupiter: An analysis of the
Voyager radio occultation measurements. {\it J. Geophys.  Res.} {\bf 86},
8721--8727.

\bib{Lindal, G.F., Sweetnam, D.N., Eshleman, V.R.} 1985. The atmosphere of
Saturn: an analysis of the Voyager radio occultation measurements with
Voyager 2. {\it J. Geophys. Res.} {\bf 92}, 14987--15001. 

\bib{Lindal, G.F.} 1992. The atmosphere of Neptune: An analysis of
radio-occultations data acquired with Voyager 2. {\it
Astron. Journal} {\bf 103}, 967--982.

\bib{Lissauer, J.J.} 1993. Planet formation. {\it
Annu. Rev. Astron. Astrophys.} {\bf 31}, 129--174.

\bib{Mahaffy, P.R, Donahue, T.M., Atreya, S.K., Owen, T.C., Niemann,
H.B.} 1998. Galileo probe measurements of D/H and $^3$He/$^4$He in Jupiter's
atmosphere. {\it Space Sci. Rev.} {\bf 84}, 251--263.

\bib{Marcy, G.W., Butler, R.P.} 1998. Detection of extrasolar giant
planets. {\it Ann. Rev. Astron. Astrophys.} {\bf 36}, 57--98.

\bib{Marley, M.S., Saumon, D., Guillot, T., Freedman, R.S., Hubbard, W.B.,
Burrows, A. \& Lunine, J.I.} 1996. On the nature of the brown dwarf
Gliese 229 B. {\it Science} {\bf 272}, 1919--1921.

\bib{Mayor, M., Queloz, D.} 1996. A Jupiter-mass companion to a
solar-type star. {\it Nature} {\bf 378}, 355--359.

\bib{Nellis, W.J., Louis, A.A., Ashcroft, N.W.} 1998. Metallization of
fluid hydrogen. {\it Phil. Trans. R. Soc. Lond. A} {\bf 356}, 119-138.

\bib{Nellis, W.J., Weir, S.T., Mitchell, A.C.} 1999. Minimum metallic
conductivity of fluid hydrogen at 140 GPa (1.4 Mbar). {\it
Phys. Rev. B} {\bf 59}, 3434--3449.

\bib{Niemann, H.B., Atreya, S.K., Carignan, G.R., Donahue, T.M.,
Haberman, J.A., Harpold, D.N., Hartle, R.E., Hunten, D.M., Kaspzrak,
W.T., Mahaffy, P.R., Owen, T.C., Way, S.H.} 1998. The composition of
the jovian atmosphere as determined by the Galileo probe mass
spectrometer. {\it \JGR} {\bf 103}, 22831--22846.

\bib{Pfaffenzeller, O., Hohl, D., Ballone, P.} 1995.
Miscibility of hydrogen and helium under astrophysical conditions,
{\it Phys. Rev. Lett.} {\bf 74}, 2599--2602.

\bib{Pollack, J.B., Hubickyj, O., Bodenheimer, P., Lissauer, J.J.,
Podolak, M., Greenzweig, Y.} 1996. {\it Icarus} {\bf 124}, 62--85.

\bib{Roos-Serote M., Drossart, P., Encrenaz, T., Lellouch, E.,
Carlson, R.W., Baines, K.H., Kamp, L., Mahlman, R., Orton, G.S.,
Calcutt, S., Irwin, P., Taylor, F., Weir, A.} 1998. Analysis of Jupiter
north equatorial belt hot spots in the 4-5 $\mu$m range from
Galileo/near-infrared mapping spectrometer observations: Measurements
of cloud opacity, water and ammonia. {\it \JGR} {\bf 103}, 23023--23042.

\bib{Roulston, M.S., Stevenson, D.J.} 1995. Prediction of
neon depletion in Jupiter's atmosphere {\it EOS} {\bf 76}, 343 [abstract].

\bib{Salpeter, E.E.} 1973. On convection and gravitational layering in
Jupiter and in stars of low mass. {\it Astrophys. J. Lett.} {\bf 181},
L183--L186.

\bib{Saumon D., Chabrier, G., Van Horn, H.M.} 1995. An equation
of state for low-mass stars and giant planets. {\it
Astrophys. J. Suppl. Ser.} {\bf 99}, 713--741. 

\bib{Saumon, D., Chabrier, G., Wagner, D.J., Xie, X.} 1999. Probing
the inner secrets of brown dwarfs and giant planets. preprint.

\bib{Seiff, A., Kirk, D.B., Knight, T.C.D., Young, R.E., Mihalov, J.D.,
Young, L.A., Milos, F.S., Schubert, G., Blanchard, R.C, Atkinson, D.}
1998. Thermal structure of Jupiter's atmosphere near the
edge of a 5-$\mu$m hot spot in the north equatorial belt. {\it \JGR}
{\bf 103}, 22857--22890.

\bib{Showman, A.P., Ingersoll, A.P.} 1998. Interpretation of Galileo
probe data and implications for Jupiter's dry downdrafts. {\it Icarus}
{\bf 132}, 205--220. 

\bib{Stevenson, D.J.} 1982. Interiors of the giant planets. {\it
Annu. Rev. Earth Planet. Sci} {\bf 10}, 257--295.

\bib{Stevenson, D.J.} 1985. Cosmochemistry and structure of the
giant planets and their satellites. {\it Icarus} {\bf 62}, 4--15.

\bib{Stevenson, D.J., Salpeter, E.E.} 1977. The phase diagram
and transport properties for hydrogen-helium fluid planets. 
{\it \ApJ\ Suppl.} {\bf 35}, 221--237. 

\bib{Thompson, S.L.} 1990. ANEOS--Analytic equations of state for
shock physics codes. Sandia Natl. Lab. Doc. SAND89-2951.

\bib{Weir, S.T., Mitchell, A.C., Nellis, W.J.} 1996. Metallization
of fluid molecular-hydrogen at 140 GPA (1.4 Mbar). {\it
Phys. Rev. Lett.} {\bf 76}, 1860-1863.

\bib{von Zahn, U., Hunten, D.M., Lehmacher, G.} 1998. Helium in
Jupiter's atmosphere: results from the Galileo probe helium
interferometer experiment. {\it J. Geophys. Res.} {\bf 103},
22815--22830. 

\bib{Zhang, K., Schubert, G.} 1996. Penetrative convection and zonal
flow on Jupiter. {\it Science} {\bf 273}, 941--943.

\bib {Zharkov V.N., Trubitsyn V.P.} 1974. Determination of the
equation of state of the molecular envelopes of Jupiter and Saturn
from their gravitational moments. {\it Icarus} {\bf 21}, 152--156.

\bib {Zharkov V.N., Trubitsyn V.P.} 1978. {\it Physics of planetary
interiors}, W.B. Hubbard ed., Pachart Press, Tucson.

\bib{Zharkov, V.N., Gudkova, T.V.} 1992. Modern models of giant
planets. In {\it High Pressure Research: Application to Earth and
Planetary Sciences} (Y. Syono and M.H. Manghgnani, Eds.),
pp. 393--401.

\newpage

\newpage

\newpage

\newpage

\newpage

\newpage

\newpage

\newpage

\newpage

\newpage

\end{document}